\documentstyle[12pt,epsfig]{article}

\newcommand{\lsim}{\mathrel{\lower4pt\hbox{$\sim$}}
\hskip-12.5pt\raise1.6pt\hbox{$<$}\;}

\newcommand{\gsim}{\mathrel{\lower4pt\hbox{$\sim$}}
\hskip-12.5pt\raise1.6pt\hbox{$>$}\;}

\def\qwe{\zeta}

\begin{document}
\baselineskip 18pt plus 2pt

\noindent \hspace*{10cm}UCRHEP-T200\\
\noindent \hspace*{10cm}August 1997\\

\begin{center}
{\bf Implications of a $W^+W^- (ZZ) - {\rm Higgs}-t \bar c$ Interaction for 
$e^+e^-\to t\bar c\nu_e\bar\nu_e$, $t\bar
ce^+e^-$, $t \bar c Z$ and for  $t\to cW^+W^-$, $cZZ$ in a Two Higgs Doublet 
Model}

\bigskip\bigskip

S. Bar-Shalom$^a$, G. Eilam$^b$, A. Soni$^c$ and J. Wudka$^a$

$^a$ Physics Dept., University of California, Riverside CA 92521, USA.\\
$^b$ Physics Dept., Technion-Israel Inst.\ of Tech., Haifa 32000, Israel.\\
$^c$ Physics Dept., Brookhaven Nat.\ Lab., Upton NY 11973, USA.

\date{\today}
\end{center}

\abstract{
\begin{quote}
{\bf Abstract}: The Standard Model 
with one extra Higgs doublet may give rise to 
enhanced {\it tree-level} 
flavor-changing-scalar coupling of a neutral Higgs to a pair of top-charm
quarks. This coupling may drive a large {\it tree-level} effective
$W^+W^-(ZZ)-{\rm Higgs}-t \bar c$ interaction. As a result we find that the
reactions $e^+e^-\to t\bar c\nu_e\bar\nu_e$, $t\bar
ce^+e^-$, $t \bar c Z$ and the two rare top decays $t\to cW^+W^-$, $t\to
cZZ$ become very sensitive probes of such an effective interaction. The most
promising ones, $e^+e^-\to t\bar c\nu_e\bar\nu_e$, $t\bar
ce^+e^-$, may yield several hundreds and up to thousands of such
events at the Next Linear Collider with a center of mass energy of
$\sqrt{s}=0.5$--2 TeV if the mass of the lightest neutral Higgs is a few
hundred GeV\null. The rare decays $t\to cW^+W^-$ and $t\to cZZ$ may 
be accessible at the LHC if the mass of the lightest neutral Higgs lies in the
narrow window $150~{\rm GeV} \lsim m_h \lsim 200$ GeV\null.
\end{quote}
}

\newpage

\section{Introduction}

Understanding the nature of the scalar sector, which still remains one of the
great mysteries in electroweak theories,
and searching for flavor-changing (FC) currents are clearly important goals
of the next generation of high energy colliders~\cite{nlc}.

Although the Standard Model (SM) with only one scalar doublet is in good
agreement with existing data, it is still useful to examine consequences
of simple extensions of the SM\null. Indeed, the simplest possible extension
of the scalar 
potential, which contains two Higgs doublets, exhibits rich new
phenomena. In particular it may give rise to new tree-level FC couplings of a
spin 0 particle with fermions~\cite{luke}.

In the SM there are no tree-level flavor-changing-neutral-currents (FCNC).
At the one loop level, FC transitions involving external up
quarks are much more suppressed than those involving external down quarks. 
The effects
for the up quarks are driven by virtual exchanges of down quarks for which
the GIM 
mechanism is much more effective since the mass splitting between the
down quarks is a lot less than amongst the charge 2/3 quarks.
Therefore, the search for large signatures of FCNC involving the up quarks is
extremely important as it may serve as a unique test of the SM\null.
As is well known, though there are stringent experimental constraints against
the existence of tree level flavor-changing-scalar (FCS) transitions
involving the light quarks~\cite{weinb,sher1,soni1}, analogous constraints
involving the top quark are essentially non-existent.

As mentioned above, a mild extension of the SM in which one extra
scalar doublet is added, allows for large, tree-level FCS
interactions~\cite{luke}. These are often forbidden by the imposition of
an {\it ad-hoc} symmetry~\cite{weinb}; if this symmetry is not imposed,
however,
one arrives at a version of the two-Higgs doublet model (2HDM) wherein
the up and down-type quarks are allowed simultaneously to couple to more
than one scalar doublet~\cite{luke} leading to tree-level FC vertices. 
In the context of such new interactions, the severe experimental constraints
involving FC couplings of the light quarks can be satisfied by requiring
that FCS interactions are proportional to the square root of masses of the
fermions participating at the vertex~\cite{sher1}. A specific realization
of these ideas, 
the Cheng-Sher Ansatz (CSA), assumes that the FC coupling of a scalar to top
and up (charm) quark is proportional to  $\sqrt{m_t m_u}/m_W$ (or $\sqrt{m_tm_c}
/ m_W$). In this scenario the large top mass makes it much more susceptible
to FC transitions. This possibility has led various authors to stress the
importance of searching for tree-level FCS interactions involving the top-quark,
especially the top-charm ones [2,5--9].
Our study indicates that experimental
investigations of the reactions $e^++e^- \to t\bar c\nu_e\bar\nu_e;~\bar
tc\nu_e\bar\nu_e;~t\bar ce^+e^-;~\bar tce^+e^-;~Zt \bar c;~Z \bar t c$ and
of the rare top decays $t \to W^+W^-c;~ZZc$ will be very useful in this regard.

The paper is organized as follows: in section 2 we briefly describe the key
features of a 2HDM with tree-level FC couplings, often called Model~III\null.
The 
possibility of producing $t \bar c$ pairs via $WW$ and $ZZ$ fusion in the next
linear collider (NLC) is investigated in section 3. In section 4 we discuss
the reaction $e^+ e^- \to Zt \bar c$. In section 5 we examine the two rare
top decays $t \to W^+W^-c$ and $t \to ZZc$ and in section 6 we summarize
our results and make some parting comments.\\

\section{2HDM With Tree-Level FC Couplings (Model III)}

In a most general version of the 2HDM (which allows tree-level FCS couplings)
one can always choose a basis of scalar fields where only one doublet
acquires a vacuum expectation 
value (VEV) (for a brief review see~\cite{soni1}):
%
\begin{eqnarray}
\langle \phi_1^0 \rangle =\frac{v}{\sqrt 2}~,~\langle \phi_2^0 \rangle =0
\label{vev}~.
\end{eqnarray}
We refer to this type of a 2HDM as Model III\null.

With this choice $\phi_1$ corresponds to the usual SM scalar doublet and all
the new FC couplings are associated with the $\phi_2$ doublet.
The spectrum of the scalar sector then consists of a charged scalar and its
conjugate $H^{\pm}$, and three neutral Higgs particles which
we will denote by $h,H$ (the scalar mass eigenstates) and $A$ (the
pseudoscalar mass eigenstate). In terms of the original doublets one has:
\begin{eqnarray}
H&=& \sqrt 2 \left[ \left({\rm Re}\phi_1^0 -v \right) \cos {\tilde {\alpha}} +
{\rm Re}\phi_2^0 \sin {\tilde {\alpha}} \right] ~,\nonumber\\
h&=& \sqrt 2 \left[ -\left({\rm Re}\phi_1^0 -v \right) \sin {\tilde {\alpha}} +
{\rm Re}\phi_2^0 \cos {\tilde {\alpha}} \right] \label{spectrum}~,\\
A&=& \sqrt 2 \left(-{\rm Im} \phi_2^0 \right)~. \nonumber
\end{eqnarray}
The masses of the neutral and charged Higgs bosons as well as the mixing angle
${\tilde {\alpha}}$ are free parameters of the model.\footnote{We use $ \tilde
\alpha $ instead of $ \alpha $ to avoid confusion with the fine-structure
constant.} The pseudoscalar $A$ which does not couple to gauge bosons and
the charged Higgs particles of the model do not play any role in our reactions and
therefore their masses are not relevant for the present analysis. 

Although with the above basis for Model~III, in which $<\phi_2^0>=0$
at the tree-level, introducing large splitting between the masses of the
two Higgs particles $h$ and $H$ (in some cases we will take $m_H-m_h > 500$
GeV) can become slightly unnatural for large values of ${\tilde {\alpha}}$,
this is not the case in a more general flavor-changing 2HDM where both doublets
can acquire a non-vanishing VEV\null. In that more general case, $\tan\beta
\equiv v_2/v_1$ appears as an additional free parameter of the model. 
Adopting $\tan\beta \neq 0$ will not affect our predictions in this paper,
while, due to the presence of this additional free parameter $\tan\beta$,
large values of ${\tilde {\alpha}}$ can be  accommodated without much difficulty
in this
framework regardless of the degree of splitting between the two Higgs masses.
Note also that $\tan\beta$ will not enter the FC couplings of a neutral Higgs
to fermions as those are governed by the couplings $\lambda_{ij}$ to be defined
below.   We therefore wish to emphasize that we are not trying to advocate the
existence of the above particularly simple realization of a FC 2HDM where
one of the Higgs doublets does not acquire a VEV, instead, for its simplicity,
we are using it as an illustrative scenario
to estimate the size of a possible FC  effect in our reactions. Thus, in
what follows, we will always choose the mass of the lighter Higgs, $h$, to
be in the range $50~{\rm GeV} \lsim m_h \lsim 1~{\rm TeV}$
while, in most instances, we will
set the heavy Higgs ($H$) mass to be $m_H=1$ TeV independent of the
choice of mixing angle ${\tilde {\alpha}}$.\footnote{Note that the onset 
of a strongly interacting Higgs sector corresponds to the breakdown of tree-level
unitarity and  also to the condition that the radiative corrections  to the
Higgs mass are of  order 100\%, i.e., $\delta m_H \approx m_H$. Much like 
in the SM case, this will occur when $m_H \sim 4 \pi v \sim 3$ TeV\null. 
Therefore, although taking $m_H=1$ TeV is somewhat  close to the above limit,
still, it is unlikely to enter the strongly interacting Higgs domain.}

The FC part of the Yukawa Lagrangian in Model~III is given
by~\cite{luke,soni1}:
\begin{eqnarray}
{\cal L}_Y^{FC}=\xi_{ij}^U {\bar Q}_{i,L} {\tilde \phi}_2 U_{j,R} +
\xi_{ij}^D {\bar Q}_{i,L} \phi_2 D_{j,R} + h.c. \label{lyfc}~,
\end{eqnarray}

\noindent where $\phi_2$ denotes the second scalar doublet, ${\tilde \phi}_2 
\equiv i \tau_2 \phi_2$, $Q$ stands for the quark doublets, and $U$ 
and $D$ for charge 2/3 and (-1/3) quarks singlets; $i,j=1,2,3$ 
are the generation indices and $\xi$ are $3\times 3$
matrices parameterizing the strength of FC neutral scalar vertices.
Following Cheng and Sher~\cite{sher1} we choose the
parameterization: 
\begin{eqnarray}
\xi_{ij}^{U,D}=g_W\frac{{\sqrt {m_im_j}}}{m_W} \lambda_{ij} \label{csa}~.
\end{eqnarray}
In this scenario all our ignorance regarding the FCS vertices
is contained in the couplings $\lambda_{ij}$ which are free
parameters to be deduced from experiments.
The experimental constraints on the $ \lambda_{ i j } $ 
are rather mild: for example, if $\lambda_{sd},
\lambda_{bd}$ and $\lambda_{uc}$ are kept below $\sim 0.1$, then
Model~III is compatible with the existing low energy
experimental measurements as long as the other FC couplings 
(i.e., those involving the top quark) are
not much larger than 1 \cite{soni1}. In particular, if the first
generation FC couplings are not related to the FC couplings of
the second and third generations (there is no good reason to
believe that such a relation exists) then $\lambda_{tc}=\lambda_{ct}
\sim O(1)$, or even somewhat bigger, is not ruled out\footnote{$\lambda_{tu}$
is also not well constrained from existing experiments. The Cheng
and Sher Ansatz (\ref{csa}) does, of course, imply much smaller $tu$
coupling compared to the $tc$ one due to the up-charm mass difference.} by
existing experiments~\cite{soni1}. This has major consequences on our
analysis in this paper as all the reactions investigated here scale like
$\lambda_{tc}^2$.

For simplicity, we choose $\lambda_{tc}=\lambda_{ct}=\lambda$ and we
furthermore break $\lambda$ into its real and imaginary parts,
$\lambda=\lambda_R+i\lambda_I$.
Then, within the CSA, the relevant terms of the Model~III Lagrangian
become:
\begin{eqnarray}
{\cal L}_{{\cal H}tc}&=&-\frac{g_W}{\sqrt 2} \frac{{\sqrt {m_t m_c}}}{m_W}
f_{\cal H} {\cal H} \bar t (\lambda_R+i\lambda_I \gamma_5) c  \label{htc}~,\\
{\cal L}_{{\cal H} VV}&=&-g_W m_W C_V c_{\cal H} {\cal H} g_{\mu\nu} V^{\mu}
V^{\nu} \label{hww}~,
\end{eqnarray}
where here and throughout the paper ${\cal H}=h~{\rm or}~H$ and $V=W~{\rm
or}~Z$ and\footnote{$V=W^+, W^-$ or $Z$; in most instances the appropriate choice can be
fixed by inspection. If necessary we will denote $V^1=W^+,V^2=W^-$ or
$V^1=V^2=Z$.}: 
\begin{eqnarray}
f_{h;H}&\equiv& \cos{\tilde {\alpha}};\sin{\tilde {\alpha}} \label{fhh}~,\\
c_{h;H}& \equiv& \sin{\tilde {\alpha}};-\cos{\tilde {\alpha}} \label{chh}~,\\
C_{W;Z}&\equiv&1;m_Z^2/m_W^2 \label{cwz}~.
\end{eqnarray}

The amplitude for the reaction $VV-{\cal H}-t \bar c,~\bar t c$
is proportional to $ \sin 2 \tilde \alpha $ for {\em both} $ {\cal H } =
h $ and $ H $, and will vanish for $ \tilde \alpha = 0 , \pi/2 $.
When ${\tilde {\alpha}}=\pi/4$ (i.e., equal mixing
between ${\rm Re}\phi_1^0-{\rm Re}\phi_2^0$) the $h$ and $H$ contributions 
interfere destructively and cancel out in the limit $ m_H \to m_h $. The
presence of this ``GIM-like'' cancellation reflects the fact that all 
complete calculations should include both neutral scalars. The maximum
of the cross section is not reached at $ \tilde \alpha = \pi /4 $ since
the scalar widths also depend on this parameter.

We will also need the ${\cal H} t \bar t$ couplings within Model~III:
\begin{eqnarray}
{\cal L}_{{\cal H} tt} = -\frac{g_W}{\sqrt 2} \frac{m_t}{m_W}
{\cal H} \bar t \left( a_{\cal H} + ib_{\cal H} \gamma_5 \right) t
\label{htt}~,
\end{eqnarray}
where
\begin{eqnarray}
&&a_h=-\frac{1}{\sqrt 2} \sin{\tilde {\alpha}} + \cos{\tilde {\alpha}}
\lambda_R~,~b_h=\cos{\tilde {\alpha}} \lambda_I \label{ah1bh1}~,\\
&&a_H=\frac{1}{\sqrt 2} \cos{\tilde {\alpha}} + \sin{\tilde {\alpha}}
\lambda_R~,~b_H=\sin{\tilde {\alpha}} \lambda_I \label{ah2bh2}~,
\end{eqnarray}
\noindent and for simplicity we set
$\lambda_{tt}=\lambda_{tc}=\lambda$.

\section{$t \bar c$ Production Through Vector-Boson Fusion}

In this section we consider the reactions (see Fig.~1):
\begin{eqnarray}
e^+e^- \to t \bar c \nu_e {\bar \nu_e};~ \bar t c \nu_e {\bar \nu_e}~,~
e^+e^- \to t \bar c e^+ e^-;~\bar t c e^+ e^- \label{eetcnn}~,
\end{eqnarray}
occurring via $W^+ W^-$ or $ZZ$ fusion, which should
be accessible to the Next generation of $e^+$-$e^-$ Linear
Colliders (NLC) currently being envisaged \cite{nlc}. We will
see that these processes are very sensitive to FC currents \cite{hep9703221}.

An extremely interesting feature of the
reactions in (\ref{eetcnn}) is that at c.m.\ energies of TeV and
above, the corresponding cross-sections can be much larger
than the ones for the simple $s$-channel reactions in Model~III: $e^+e^- \to t \bar c$
(see \cite{atwood1}) and $e^+e^- \to {\cal H} A \to t \bar c f \bar f;~t \bar t c \bar c$
(see \cite{houlin}).   For example, we find that $\sigma^{\nu \nu tc}
\equiv \sigma(e^+e^- \to t \bar c \nu_e {\bar {\nu}}_e + \bar t c \nu_e
{\bar {\nu}}_e)$ is about two orders of magnitude larger than
$\sigma(e^+e^- \to t \bar c + \bar t c)$ over a large region of parameter
space, while $\sigma^{eetc} \equiv \sigma(e^+e^- \to t \bar
c e^+ e^- + \bar t c e^+ e^-)$ is about one order of magnitude bigger
than $\sigma(e^+e^- \to t \bar c + \bar t c)$. The crucial difference (and
therefore interesting) feature of the $VV$ fusion reactions is that, being
a t-channel fusion process, the corresponding cross-sections {\it grow} with
the c.m.\ energy of the collider. On the other hand, the ``simple'' $s$-channel
reactions mentioned above {\it drop} like $1/s$.      
 Thus, even if no $t \bar c$ events are detected at
$\sqrt s =500$ GeV via $e^+e^- \to t \bar c;~t \bar c f \bar f;~t \bar t
c \bar c$, there is still a strong motivation to look for a 
signature of (\ref{eetcnn}) especially at somewhat higher energies.

In exploring the reactions $e^+e^-\to t\bar c\nu_e\bar\nu_e$, $t\bar c
e^+e^-$ we will use the effective vector boson approximation (EVBA)
\cite{cahn}. Recall that this is the analog of the equivalent photon
approximation in QED which allows the colliding $W$'s or $Z$'s to be
treated as on-shell particles. The salient features of the reactions in
(\ref{eetcnn}) are then well approximated by the simpler fusion
reactions:
\begin{equation}
W^+W^-, Z Z \to t \bar c, \bar tc \label{vvtc}~.
\end{equation}
The corresponding cross sections for the reactions in
(\ref{eetcnn}) can then be calculated by folding in the distribution
functions $f^V_{h_V}$, for a vector boson $V$ ($W$ or $Z$) with
helicity $h_V$.

The EVBA has been extensively studied in the production of a $t\bar t$
pair~\cite{wwtt}. There is, however, a significant difference between fusion
reactions leading to  a $t\bar c$ final state, due primarily to the appreciable
difference in the threshold of the two-reactions (which, in turn, is due
to $m_t\gg m_c$). This has two consequences:
\begin{enumerate}
\item For $t\bar c$ the
vector-boson energy fraction, $x=\sqrt{\hat s/s}$ (as usual $\hat s$ is the
c.m.\ energy squared in the $VV$ c.m.\ frame and $s$ the corresponding
quantity in the $e^+e^-$
c.m.\ frame), can drop below $ x = 0.05 $ near threshold, for $\sqrt{s}\gsim
800$ GeV\null. In this small-$x$ range the distribution
functions are overestimated within the leading log approximation
\cite{wwtt,wapprox1}. We will therefore use the distribution functions
which retain higher orders in $m^2_V/s$, as given, for example, by
Johnson {\it et. al.}~\cite{wapprox1}.
\item For large $\sqrt{\hat s}$, the longitudinal
polarization vector of $V$ can be approximated by $\epsilon^\mu_0(k)
\simeq k^\mu/m_V +{\cal O}(m_V/\sqrt{\hat s})$. In the production of a pair of
heavy fermions (such as $t\bar t$) through $VV$ fusion, the term
$k^\mu/m_V$ gives rise to a contribution 
proportional to  $(m_t/m_V)^4$ in the cross section; the
subleading contributions, generated by the
${\cal O}(m_V/\sqrt{\hat s})$ remainder in $\epsilon^\mu_0(k)$,
are suppressed by a factor of $\sim m^2_t/\hat s$. Thus
$\hat\sigma (VV\to t\bar t)$ is well approximated by taking only the
longitudinal polarized $V$'s at the parton level reaction and assuming
that $\hat s\gg m^2_V$~\cite{wwtt,wapprox1}. In contrast, the
approximation $\epsilon^\mu_0(k) \simeq k^\mu/m_V $ does
not necessarily hold for the reaction $VV\to t\bar c$ for which
$m^2_V/\hat s \approx m^2_V/m^2_t$ near threshold. In particular, we
will show below that the cross-section for the reaction $VV\to {\cal
H}\to t\bar c$ scales
like $|\epsilon^{V^1}_{h_{V^1}} \cdot \epsilon^{V^2}_{h_{V^2}}|^2$.
Thus, not only is the $(m_t/m_V)^4$ factor absent, but
the contribution from the transversely polarized $V$'s is
comparable to that of the longitudinal $V$'s near threshold. We will therefore
include all polarizations for the vector bosons in our calculation of
$\hat\sigma(VV\to {\cal H}\to t\bar c)$.
\end{enumerate}

It is interesting to note that while at tree-level, $\sigma^{eetc}=0$ in the
SM, the parton level reaction $W^+W^- \to t \bar c$ can proceed at tree-level,
via diagram a in Fig.~1. Note that the corresponding cross section
is proportional to $(m_t/m_W)^4$ to leading order, and the usual replacement
 $\epsilon^{{\mu}}_0(k) \to k^{{\mu}}/m_W$ is appropriate. For collision of
longitudinal $W$'s, $W^+_L W^-_L \to t \bar c$, within the SM, we obtain:
\begin{eqnarray}
{\hat {\sigma}}_{\rm SM}=\frac{N_c \pi \alpha^2}{4 s^4_W {\hat s}^2}
 \left(\frac{m_t}{m_W}\right)^4 \sum_{i,j=1}^3 V_{ti}V_{tj}^*V_{ci}V_{cj}^*
\left\{ \left( \frac{1}{\Delta_t} -1 \right) I_{ij}^2 + \left( 2
-\frac{1}{\Delta_t} \right) \frac{I_{ij}^3}{m_t^2} - \frac{I_{ij}^4}{m_t^4}
\right\} \label{sig0sm} ~,
\end{eqnarray}
where $i,j$ are family indices, $s_W \equiv \sin\theta_W$ and
$N_c=3$ is the color factor. $\Delta_t \equiv m_t^2/\hat s$ and
$I_{ij}^k$ are the two body phase-space integrals:

\begin{eqnarray}
I_{ij}^k \equiv \int_{m_t^2- \hat s}^{0} \frac{x^k
dx}{(x-m_{d_i}^2)(x-m_{d_j}^2)} \label{phspint}~.
\end{eqnarray}

\noindent In (\ref{sig0sm}) we have set $m_c=0$ , however the three down quarks
masses must be kept non-zero as the unitarity of the Cabibbo-Kobayashi-Maskawa
(CKM) matrix implies that ${\hat {\sigma}}_{\rm SM}=0$ when $m_d=m_s=m_b$ (in
particular, when $m_d=m_s=m_b=0$). Numerically, $ \hat \sigma_{\rm SM} $
is found to be too small to be of experimental
relevance as it suffers from a severe CKM suppression: $\sigma^{\nu \nu
tc}_{\rm SM} \equiv \sigma_{\rm SM}(e^+ e^- \to t \bar c \nu_e \bar{\nu_e}
+ \bar t c \nu_e \bar{\nu_e}) \approx 10^{-5}-10^{-4}$ (fb) 
for $\sqrt s = 0.5 - 2$ TeV. We will hence forward neglect the SM contribution.

This is, therefore, a remarkable situation which allows for a unique test of the SM and, in
particular of the SM's GIM mechanism.
Even a very small number of  $t \bar c \nu_e {\bar {\nu}}_e$ and/or
$t \bar c e^+e^-$ detected at a NLC running with a yearly integrated luminosity
of ${\cal L} \gsim 10^2$ [fb]$^{-1}$~\cite{nlc}, will unmistakably indicate
new FC dynamics beyond the SM\null. 
In Model~III event numbers in the
range of a few${}\times (10^2-10^3)$ for
$t \bar c \nu_e {\bar {\nu}}_e$, and a few${}\times (10^1-10^2)$ for
$t \bar c e^+e^-$ are easily  possible within the existing experimental
constraints. 

For Model~III, $VV \to t \bar c$ proceeds at tree-level
via the $\hat s$-channel neutral Higgs exchange of diagram b in Fig.~1.
Neglecting the SM diagram, the corresponding parton-level cross-section
${\hat \sigma}_V \equiv {\hat \sigma}(V^1_{h_{V^1}}V^2_{h_{V^2}} \to t \bar c)$
is given by~\cite{hep9703221}:
\begin{eqnarray}
{\hat \sigma}_V &=
\frac{\left( \sin 2{\tilde \alpha } \right)^2 N_c \pi \alpha^2}{4 {\hat s}^2
\beta_V s^4_W}  \left(\frac{m_V}{m_W}\right)^4 
| \epsilon^{V^1}_{h_{V^1}} \cdot \epsilon^{V^2}_{h_{V^2}}|^2  
|\Pi_h - \Pi_H |^2 \times 
\nonumber \\
&  m_t m_c \sqrt {a_+ a_-} ( a_+ \lambda_R^2 + a_- \lambda_I^2 )    
\label{vvhtc}~,
\end{eqnarray}
\noindent where:
\begin{eqnarray}
a_\pm = \hat s - ( m_t \pm m_c )^2~~,~~ 
\beta_\ell \equiv \sqrt {1-4 m_{\ell}^2/\hat s}~,
\end{eqnarray}
\noindent and:
\begin{eqnarray}
\Pi_{\cal H} = \frac{1}{\left( \hat s - m_{\cal H}^2 +i m_{\cal H} 
{\Gamma_{\cal H}} \right)}~.
\end{eqnarray}
%
%
%
Given the couplings of Model~III, $\Gamma_{\cal H}$ (the width of ${\cal H}$)
can be readily calculated~\cite{soni3}. The leading
decay rates in this model are ${\cal H} \to b\bar b ,t \bar t ,ZZ,W^+W^-$
and $t \bar c, c \bar t$. When kinematically allowed, we include all these
contributions in calculating the above cross-sections. 
For definiteness, we will present our numerical results for
${\tilde {\alpha}}=\pi/4$.\footnote{As will be shown later, the $VV$ fusion
cross-sections in (\ref{eetcnn})
reach their maxima at $ \tilde \alpha \simeq \pi/6 $ which is larger by a
factor of $ \sim 1.5 $ than their value at $ \tilde \alpha = \pi/4 $;
as indicated previously the cross sections vanish when $ \tilde \alpha 
= 0 , \pi /2 $.} We will also ignore CP violation and take $\lambda_I=0$ and
$\lambda=\lambda_R$.  In calculating the cross sections we first vary
the mass of the lighter scalar $h$ in the range $100~{\rm GeV}<m_h<1~{\rm
TeV}$, while holding fixed the mass of the heavy scalar $H$ at $m_H=1$ TeV\null.
We will later discuss the case $m_h \sim m_H$.

Due to the orthogonality properties of the $V^1$ and $V^2$
polarization vectors, there is no interference between the transverse
and the longitudinal polarizations. Note that
$| \epsilon^{V^1}_{\pm} \cdot \epsilon^{V^2}_{\mp}|^2=0$,
$| \epsilon^{V^1}_{\pm} \cdot \epsilon^{V^2}_{\pm}|^2=1$, and $|
\epsilon^{V^1}_0 \cdot \epsilon^{V^2}_0|^2 = (1+\beta_V^2)^2/(1-\beta_V^2)^2$
which grows with $\hat s$. However, we can see from (\ref{vvhtc}),
that ${\hat {\sigma}}_V(m_{\cal H}^2/\hat s \to 0) \to 0$ ensuring unitarity of
the hard cross-section.
In general, the transverse distribution functions are bigger than the
longitudinal ones for $x$ $\gsim 0.1$~\cite{wwtt,wapprox1}. Therefore,
the relative smallness of the transverse hard cross-section compared
to the longitudinal one is partly compensated for in the full cross-section.
In particular, we find that the contribution from the transversely polarized
$W$'s($Z$'s) constitutes up to 25\% (35\%) of the corresponding
total cross-section $\sigma^{\nu \nu tc}$($\sigma^{eetc}$).

It is evident from (\ref{vvhtc}) that ${\hat \sigma}_W \to {\hat \sigma}_Z$
for $m_W \to m_Z$.
The main difference between $\sigma^{\nu \nu tc}$ and $\sigma^{eetc}$
then arises from the dissimilarity between the distribution functions
for $W$ and $Z$ bosons. In particular, disregarding the subleading
transverse parts of the $WW$ and the $ZZ$ cross-sections, the relative
strength between the $W$ and the $Z$ longitudinal distribution functions
is given by \cite{wapprox1}:
\begin{eqnarray}
f^Z_0=\frac{2}{c^2_W}\left( 2s^4_W - s^2_W + \frac{1}{4} \right) f^W_0 \approx
\frac{1}{3} f^W_0 \label{distfun}~.
\end{eqnarray}
Therefore, since the dominant contributions to the cross-sections $\sigma^{\nu
\nu tc}$ and $\sigma^{eetc}$ are produced by longitudinal $W$'s
and $Z$'s, $\sigma^{eetc}$ is expected to be smaller by
about one order of magnitude than $\sigma^{\nu \nu tc}$, which is
indeed what we find. We will thus only present
numerical results for $\sigma^{\nu \nu tc}$, keeping in mind
that $\sigma^{eetc}$ exhibits the same behavior though suppressed by an
order of magnitude.

Fig.~2 shows the dependence of the scaled cross-section
$\sigma^{\nu \nu tc}/\lambda^2$ on the mass of the light Higgs $m_h$ for
four values of $s$.\footnote{The scaled cross-section,
$\sigma^{\nu \nu tc}/ \lambda^2$, has a residual mild dependence on
$\lambda$ through its dependence on  $\Gamma_h$.} The cross-section peaks at
$m_h \simeq 250$ GeV and drops as the mass of the light Higgs approaches
that of the heavy Higgs due to the ``GIM-like'' cancellation present
in the scalar sector (which is only partly effective when $ \tilde
\alpha \not= \pi/4 $). Nonetheless, as will be shown below, $\sigma^{\nu \nu tc}/ \lambda^2$ can stay at the fb level even for $m_h = m_H$.
When $\sqrt s =2$ TeV the cross-section is about 5 fb for
$\lambda=1$ and $m_h \approx 250$ GeV\null.\footnote{The cross-section
is $\propto \lambda^2$ so that even a moderate change of $\lambda$, say by a
factor of three, can increase or decrease the cross-section by one order of
magnitude.}
It is therefore evident from Fig.~2 that at an NLC running at energies of
$\sqrt s \gsim 1$ TeV and an integrated luminosity of the order
of ${\cal L} \gsim 10^2$ [fb]$^{-1}$, Model~III (with $\lambda=1$) predicts
hundreds and
up to thousands of $t \bar c \nu_e {\bar {\nu}}_e$ events and several tens
to hundreds of $t \bar c e^+ e^-$ events. For
example, with $\sqrt s=1.5$ TeV, ${\cal L}= 500$ [fb]$^{-1}$~\cite{nlc}, and
$m_h \approx 250$ GeV, $\lambda=1$, the cross-section
$\sigma^{\nu \nu tc}$($\sigma^{eetc}$) would yield about 2000(200)
such events. Note also that even with $m_h \approx 500$ GeV,
this projected luminosity will still yield hundreds of
$t \bar c \nu_e {\bar {\nu}}_e$ events and tens of
$t \bar c e^+ e^-$ events at $\sqrt s=1.5$ TeV. The corresponding SM prediction
yields, as shown above, essentially zero events.

The choice  $\tilde {\alpha}= \pi/4$ is special in the sense that for
this value the GIM-like cancellation mentioned above is most effective, however,
it does not correspond to the maximum of the production rates. In Fig.~3
we show the dependence of 
$\sigma^{\nu \nu tc}/\lambda^2$ on $(\sin\tilde {\alpha})^2$ for $m_h=250$
GeV, $\sqrt s = 1$ TeV and for two possible values of $m_H$, $m_H=250$ GeV
and $m_H=1$ TeV\null. The same behavior is observed for any value of $s$ in
the range 0.5--2 TeV\null. We see that for $m_H=1$ TeV, which
represents the case of large splitting between the two neutral Higgs particles,
 $\sigma^{\nu \nu tc}( \pi/ 14 \lsim \tilde {\alpha} \lsim \pi/4) >
\sigma^{\nu \nu tc}(\tilde {\alpha}= \pi/4)$. Moreover, even for $(m_H-m_h)
\approx
0$, $\sigma^{\nu \nu tc} \gsim 1$ fb is still possible for $ 0.02 \lsim
(\sin\tilde {\alpha})^2 \lsim 0.22$ and $ 0.78 \lsim (\sin\tilde {\alpha})^2
\lsim 0.98$. In fact, our analysis shows that, with moderate restrictions on
$\tilde {\alpha}$, $\sigma^{\nu \nu tc}$
remains well above the fb level for $\sqrt s \gsim 1$ TeV
as long as one of the neutral Higgs particles is kept within
$200 ~{\rm GeV} \lsim m_{\cal H} \lsim 400~{\rm GeV}$, while the mass of the
other
Higgs can take practically any value between 100 GeV -- 1000 GeV\null. Moreover,
note that as $\tilde {\alpha}$ drops below $\pi/6$ the cross-section becomes
less sensitive to the heavy Higgs mass. For example, we find that with $\tilde
{\alpha} \approx \pi/27$ (which may represent the case of a small $\tilde
{\alpha}$) and for $\sqrt s=1$ TeV, $\sigma^{\nu \nu t c} \approx 1$ fb 
regardless
of the heavy Higgs mass (i.e., $m_H=250 - 1000$ GeV) and as one goes to $\sqrt
s >1$ TeV, $\sigma^{\nu \nu t c}$ becomes even bigger. It is therefore clear
that the FC effect being investigated in this section remains very interesting
within a large portion of the free parameter space of the Higgs sector in
Model~III\null.  

Before ending this section we wish to comment further on the comparison between
the cross-section $\sigma(e^+e^- \to t \bar c)$ discussed in~\cite{atwood1} and
the $WW$ annihilation cross-section $\sigma^{\nu \nu tc}$ within Model~III\null.
To do so, for convenience, we normalize the cross sections to
the $\mu^+\mu^-$ cross-section:
\begin{eqnarray}
R^{\nu \nu tc} \equiv \frac{\sigma(e^+e^- \to t \bar c \nu_e {\bar {\nu}}_e +
\bar t c \nu_e {\bar {\nu}}_e)}{\sigma(e^+e^- \to \gamma \to \mu^+\mu^-)} ~,
\qquad
R^{tc} \equiv \frac{\sigma(e^+e^- \to t \bar c + \bar t c)}{\sigma(e^+e^- \to
\gamma \to \mu^+\mu^-)} ~.
\end{eqnarray}

\noindent Note that while $R^{\nu \nu tc}$ scales as $\lambda^2$, $R^{tc}$ is
proportional to $\lambda^4$.
It was shown in~\cite{atwood1} that $R^{tc}/\lambda^4$ can reach
$10^{-5}$ for a light Higgs mass around 200 GeV and c.m. energy of $\sqrt
s=500$ GeV. 
As the c.m.\ energy is increased $R^{tc}/\lambda^4$ stays fixed at the
$10^{-5}$ level due to the $\sim 1/s$ behavior of $\sigma(e^+e^- \to
t \bar c + \bar t c)$ with one loop FC Higgs exchanges.

In Figs.~4 and 5 we have plotted $R^{\nu \nu tc}/\lambda^2$ as a function of
$m_h$ and $\sqrt s$, respectively. We  see that for $m_h \approx 250$ GeV
and a c.m. energy of $\sqrt s =500$ GeV $R^{\nu \nu tc}/\lambda^2$ peaks at
around $10^{-3}$, two orders of magnitude above $R^{tc}/\lambda^4$.
We therefore expect the number of $t \bar c \nu_e {\bar {\nu}}_e$ events
in the NLC to be bigger by about two orders of magnitude than the number
of $t \bar c$ events.
Moreover, while the cross-section for producing a pair of $t \bar c$ sharply
drops as $\sqrt s$ is increased, the $WW$ fusion cross-section,
$\sigma^{\nu \nu t c}$ grows with $s$. In particular, Figs.~4 and 5 show
that for $200~{\rm GeV} \lsim m_h \lsim 400~{\rm GeV}$,
$R^{\nu \nu tc}/\lambda^2 \sim 10^{-2}$ for $\sqrt s \sim 1$ TeV and
$R^{\nu \nu tc}/\lambda^2 \sim 10^{-1}$ for $\sqrt s \sim 2$ TeV.

\section{ $e^+e^- \to t
\bar c \nu_e {\bar {\nu}}_e$ vs. $e^+e^- \to t \bar t \nu_e
{\bar {\nu}}_e$ and Background Considerations}

In order to give the reader a qualitative feel for the effectiveness of the $t \bar c \nu_e \bar{\nu_e}$ production rate it is instructive to compare it, in Model~III, to the production rate of the ``normal'' $e^+e^- \to t \bar t \nu_e
{\bar {\nu}}_e$. We recall that $\sigma^{\nu \nu tt} \equiv
\sigma(e^+e^- \to W^+ W^- \nu_e {\bar {\nu}}_e \to t \bar t \nu_e
{\bar {\nu}}_e)$ is dominated by collisions of two longitudinal $W$'s
at the parton level~\cite{wwtt}.
The reaction $W^+W^- \to t \bar t$ can proceed through the $t$-channel
$b$ quark exchange and the $s$-channel $\gamma,Z,h$ and $H$ exchanges
(the diagrammatic description can be found in~\cite{wwtt,ttnunudavid}).

The helicity amplitudes ${\cal A}_{\eta = \bar \eta }$,
${\cal A}_{\eta = - \bar \eta }$  ($ \eta$ and $\bar \eta $ denote
the helicities of the $t$ and $ \bar t $ quarks, respectively) for
$W^+_LW^-_L \to t \bar t$, including all the contributing diagrams,
are given by:
\begin{eqnarray}
{\cal A}_{\eta = \bar \eta } &=& \frac{\pi \alpha}{s^2_W}
{ m_t \sqrt{ \hat s } \over  m_W^2}\ 
 \left( \left[ (1 + \beta_t^2 ) \cos\theta - 2 \beta_t
\over 1 + \beta_t^2 - 2\beta_t \cos \theta \right]
- \sum_{ {\cal H} = h , H } \sqrt 2 c_{\cal H} \left( a_{\cal H}
\beta_t - i \eta b_{\cal H} \right) \Pi_{\cal H} \right) \label{aetaeta}~,\\
{\cal A }_{ \eta = - \bar \eta } &&= \frac{2 \pi \alpha}{s^2_W}
{ m_t^2 \over m_W^2} \left( \eta + \beta_t
\over 1 + \beta_t^2 - 2 \beta_t \cos \theta  \right) \sin \theta
\label{aetameta}~,
\end{eqnarray}
where $\theta$ is the c.m.\ scattering angle and $ a_{\cal H} $, $b_{\cal H}$
and $c_{\cal H}$ are given in (\ref{ah1bh1}), (\ref{ah2bh2}) and (\ref{chh}).
In the SM limit ${\tilde {\alpha}}=-\pi/4$ and $\lambda_R,\lambda_I=0$, the
hard cross-section for $W^+_LW^-_L \to t \bar t$, obtained from (\ref{aetaeta})
and (\ref{aetameta}) agrees with the one obtained by Eboli {\it et. al.}
in~\cite{wwtt}.

We give below only the ``non-standard'' parts
${\hat {\sigma}}_{hh},{\hat {\sigma}}_{HH},{\hat {\sigma}}_{hH},{\hat
{\sigma}}_{bh}$
and ${\hat {\sigma}}_{bH}$:\footnote{The
SM-like parts can be extracted from the paper by Eboli {\it et. al.}
in~\cite{wwtt} by changing
the appropriate quantum numbers of the final state fermions.}
\begin{eqnarray}
{\hat {\sigma}}_{hh}&=&{\cal G}_t (\sin {\tilde {\alpha}})^2 |\Pi_h|^2
\left(\beta_t^2 a_h^2 +b_h^2 \right) ~,\\
{\hat {\sigma}}_{HH}&=&{\cal G}_t (\cos {\tilde {\alpha}})^2 |\Pi_H|^2
\left(\beta_t^2 a_H^2 + b_H^2 \right) ~,\\
{\hat {\sigma}}_{hH}&=&-{\cal G}_t \sin 2{\tilde {\alpha}} {\rm Re}(\Pi_h
\Pi_H^*)  \left(\beta_t^2 a_ha_H + b_hb_H \right) ~,\\
{\hat {\sigma}}_{bh}&=&{\cal G}_t \sin{\tilde {\alpha}} \; a_h (1-\Delta_h)
|\Pi_h|^2  \left[ \frac{(1-\beta_t^2)^2}{2 \beta_t} \Lambda - (1+\beta_t^2)
\right] ~,\\
{\hat {\sigma}}_{bH}&=&-{\cal G}_t \cos{\tilde {\alpha}} \; a_H (1-\Delta_H)
|\Pi_H|^2  \left[ \frac{(1-\beta_t^2)^2}{2 \beta_t} \Lambda - (1+\beta_t^2)
\right] ~,
\end{eqnarray}
where $ \hat \sigma_{i j  }, \, i \not= j  $ denotes the interference
cross-section of
the $i$ and $j$ intermediate states, and:
\begin{eqnarray}
{\cal G}_t \equiv \frac{N_c \pi \alpha^2}{4 s^4_W} \frac{m_t^2}{m_W^4}
\beta_t~, \qquad \Lambda \equiv {\rm ln} \left( \frac{\beta_t+1}{\beta_t-1}
\right) ~.
\end{eqnarray}

In Fig.~6 we plot the ratio $R^{tc/tt}\equiv
\sigma^{\nu \nu tc}/\sigma^{\nu \nu tt}$ within Model~III
for $\lambda=1$,\footnote{Recall that we have assumed for simplicity
that $\lambda_{tt}=\lambda_{tc}=\lambda$.} ${\tilde {\alpha}}=\pi/4$
and $m_H=1$ TeV as a function of the light Higgs mass $m_h$ and for
$\sqrt s=0.5,1,1.5,2$ TeV\null.
$\sigma^{\nu \nu tt}$ depends very weakly on $m_h$, with a small
peak at $m_h\simeq 400$ GeV which fades as $\sqrt s$ grows.
Therefore, $R^{tc/tt}$ peaks with $\sigma^{\nu \nu tc}$ at $m_h \simeq250$
GeV\null. 
We can see from Fig.~6 that for $\sqrt s=0.5$ TeV and in the range
$200 ~{\rm GeV} \lsim m_h \lsim 400 ~{\rm GeV}$, $R^{tc/tt}>1$.
In particular, for $m_h \approx 250$ GeV, $\sigma^{\nu \nu tc}$ can become
almost two orders of magnitude larger than $\sigma^{\nu \nu tt}$.
As $\sqrt s$ grows, $R^{tc/tt}$ drops. In the range
$200 ~{\rm GeV} \lsim m_h \lsim 400 ~{\rm GeV}$, we find that for
$\sqrt s=1$ TeV, $R^{tc/tt}>0.1$, while for $\sqrt s=1.5-2$ TeV,
$0.01 \lsim R^{tc/tt} \lsim 0.1$.

The dependence of $\sigma^{\nu \nu tt}$ on $\lambda$ is significant only
near its peak (at $ m_h \sim 400 $ GeV); for
$200 ~{\rm GeV} \lsim m_h \lsim 400 ~{\rm GeV}$, where $R^{tc/tt}$
acquires its largest values, $R^{tc/tt}$ roughly scales as $\lambda^2$.
Thus, again a mild change in $\lambda$, can alter $R^{tc/tt}$ appreciably.
Hence, within Model~III, with $m_h$ in
the few-hundred GeV range, it is possible to observe
comparable production rates for the $t \bar c \nu_e {\bar {\nu}}_e$ and
$t \bar t \nu_e {\bar {\nu}}_e$ even at a NLC running at a TeV range c.m.\
energies. 

We have not done any serious study on the issue of backgrounds. 
For example,   
$\nu_e\bar\nu_e W^+W^-$ is expected to be about an order of magnitude bigger
than $\nu_e\bar\nu_e t\bar t$ and therefore could be of concern. However,
we remark that  the NLC literature suggests that detection 
of $t$ (or $\bar t$) via the main mode $t \to b q q'$ (i.e., 3-jet events)
with the constraint $m_{\rm jet1} + m_{\rm jet2} = m_W$ can be achieved with
a relatively high  efficiency \cite{frey}. The $\nu \nu W W$ cross-section
also has distinctive constraints on it that, along with the rather clean
$t$ detection, are expected to be very effective in separating it from $\nu
\nu tt$ or $\nu \nu tc$ final states. In the case of the $\nu_e \bar{\nu_e}
t \bar c$ final state, in addition to the top-quark detection via, for example,
the 3-jet mode, the other (charm) jet is rather unique and should stand out
as essentially a light quark jet, i.e., the event should look like a {\it
single} top quark event. Therefore, it will be difficult to fake a $t
\bar c$ event with a $t \bar t$ or $WW$ event.

\section{The Reaction $f \bar {f'} \to V t \bar c$}

In this section we explore the possibility of observing a signature of a
$Z t \bar c$ final state (and its conjugate one) at the NLC\null. Within
Model~III, 
the reaction $f \bar {f'} \to V t \bar c$ ($V=Z,W^+$ or $W^-$ depending on the
quantum numbers of the $f \bar {f'}$ initial state) proceeds at tree-level
via the Feynman 
diagram depicted in Fig.~7. Of course, disregarding the incoming $f \bar {f'}$
fermions, this reaction is directly related to the sub-process $VV \to {\cal H}
\to t \bar c$. We can therefore express the cross-section $\sigma ( f \bar {f'}
\to V t \bar c)$ in terms of the hard cross-section ${\hat {\sigma}}_V$ given
in (\ref{vvhtc}):
\begin{eqnarray}
\sigma ( f \bar {f'} \to V t \bar c) &=&
\frac{\alpha}{6 \pi (\sin 2 \theta_W)^2}
\Delta_V \Pi_V^2
\left\{ \left[ a_L^{f(V)} \right]^2+ \left[ a_R^{f(V)} \right]^2 \right\}
\times \nonumber\\
&& \int_1^{(\Delta_t^{-1/2} - \qwe_V)^2} \!\!\!\!\!\!dz \; \omega_1 \omega_2
\frac{\omega_1^2 +12 \Delta_V}{\omega_2^2 + 12 \qwe_V^4} ~
\sum_{h_{V^1},h_{V^2}}
\left. {\hat {\sigma}}_V \right|_{\hat s = m_t^2 z} \label{eeztc}~.
\end{eqnarray}
Here $\Delta_{\ell} \equiv m_{\ell}^2/s$ ($s$ being the c.m. energy of
the colliding $f \bar {f'}$ fermions) and $\Pi_V=(1- \Delta_V)^{-1}$.
Also $\qwe_{\ell} \equiv m_{\ell}/m_t$ and $\omega_1$, $\omega_2$ are
function of $z$ given by:
\begin{eqnarray}
\omega_1&=& \sqrt { \left(1-(\sqrt {\Delta_V} + \sqrt {\Delta_t z}) \right)
\left( 1-(\sqrt {\Delta_V} - \sqrt {\Delta_t z}) \right)} ~,\\
\omega_2 &=& z \sqrt {1- 4 z^{-1} \qwe_V^2} ~,
\end{eqnarray}
and we have defined the $Vf \bar {f'}$ interaction lagrangian as:

\begin{eqnarray}
{\cal L}_{V_{\mu}f f'} \equiv \frac{g_W}{c_W} V_{\mu} \gamma_{\mu}
\bar {f'} \left( a_L^{f(V)} L + a_R^{f(V)} R \right) f ~,
\end{eqnarray}
where $L(R)=(1 \mp \gamma_5)/2$.

The formula given in (\ref{eeztc})
is general and can be applied, for example, for calculating the sub-process
cross-sections $u \bar u , d \bar d \to Z t \bar c$ and $u \bar
d;~ \bar u d \to W^+ t \bar c;~W^- t \bar c$ relevant for
hadron colliders. Here we wish to concentrate only on the cross-section
$\sigma^{Ztc}
\equiv \sigma(e^+ e^- \to Zt \bar c + Z \bar t c)$ relevant for the NLC and
for which 
$V=Z$, $f=e^-$, $\bar {f'}=e^+$ and $a_L^{e(Z)}=1/2 - s_W^2$,
$a_R^{e(Z)}=s_W^2$. The production of a real Higgs boson and a $Z$ boson
via $e^+ e^- \to Z \to Z {\cal H}$ followed by the ${\cal H}$
decay ${\cal H} \to t \bar c$ was investigated in~\cite{houlin}.
This is of relevance whenever there is sufficient energy to produce a
real $Z {\cal H}$  pair and $m_{\cal H} > m_t+m_c$, then:
\begin{eqnarray}
\sigma(e^+e^- \to Z {\cal H} \to Z t \bar c +Z \bar t c)
\approx \sigma (e^+ e^- \to Z \to Z {\cal H}) \times {\rm Br}
\left( {\cal H} \to t \bar c + \bar t c \right) \label{realh}~.
\end{eqnarray}
Here we will extend the analysis performed in \cite{houlin} by including
both neutral Higgs particles, produced either as real or virtual particles.

In Fig.~8 we plot $\sigma^{Ztc}/ \lambda^2$ as a function of the light Higgs
mass, $m_h$, for various values of $\sqrt s $, and in Fig.~9, $\sigma^{Ztc}/
\lambda^2$ as a function of $s$ 
for various values of $m_h $ ($m_H=1$ and $\tilde {\alpha}=\pi /4$ are
kept fixed). We see that there is a significant
difference between $\sigma^{Ztc}$ and $\sigma^{\nu \nu tc},\sigma^{eetc}$; the
former drops with $s$ (as expected for an s-channel process)
while the latter increase with $s$.
Therefore, a search for a $Ztc$ signature will be most effective at lower
energies. In particular, we find that $\sigma^{Ztc}/ \lambda^2$ peaks when the
c.m.\ energy is a few tens of GeV above the threshold for producing a real $hZ$
pair. At $\sqrt s =500$ GeV and for $ 200 ~{\rm GeV} \lsim m_h \lsim 350 ~{\rm
GeV}$, $\sigma^{Ztc}/ \lambda^2 \gsim 0.2$ fb and peaks for $m_h \approx 250$
GeV at $\sim 0.6$ fb. In this range $h$ is produced on-shell and then decays to
$t \bar c$. 

Apart from the overall factor of $(\sin 2 \tilde
{\alpha})^2$ in the cross-section (from the $VV \to {\cal H} \to t \bar c$
matrix element), there is an additional strong dependence on $\tilde {\alpha}$
coming from ${\rm Br}(h \to t \bar c + \bar t c)$. This quantity also
generates a strong suppression (for $\tilde
{\alpha}= \pi /4$, ${\rm Br}(h \to t \bar c + \bar t c) \approx 10^{-2}$) since
$h$ decays mainly into $W$ pairs: 
${\rm Br}(h \to W^+W^-) \sim 1$ for $\tilde {\alpha}= \pi /4$ and $2 m_t
> m_h >2m_W$ and ${\rm Br}(h \to W^+W^-) \sim 0.7 \gg {\rm  Br}(h \to t \bar
t)$ when $m_h > 2m_t$. In contrast, within the SM ${\rm Br}(h \to W^+W^-)
\sim {\rm Br}(h \to t \bar t) \sim 0.5$ for $m_h > 2 m_t$.

Similar to the $VV$ fusion case, when there is large splitting
between the masses of the two neutral scalars (i.e., $m_H=1$ TeV),
$\sigma^{Ztc}/ \lambda^2$ is maximized for $\tilde {\alpha} \approx \pi/6$. In
Fig.~10 we plot $\sigma^{Ztc}/ \lambda^2$ as a function of $(\sin
\tilde {\alpha})^2$ for $\sqrt s = 500$ GeV, $m_h=250$ GeV and $m_H=250,1000$
GeV.\footnote{Here also, the same behavior as a function of $(\sin \tilde
{\alpha})^2$ occurs for higher energies.} As can be seen by comparing Fig.~3
with Fig.~10, $\sigma^{Ztc}$ and $\sigma^{\nu \nu t c}$ exhibit the same
dependence on $\tilde {\alpha}$ since both reactions are governed by
the $VV - {\cal H} - t c$ amplitude; we again find
that for $m_H=1$ TeV, $\sigma^{Ztc}( \pi/ 14 \lsim \tilde {\alpha} \lsim \pi/4)
> \sigma^{Ztc}(\tilde {\alpha}= \pi/4)$. When $m_H \approx m_h \approx 250$ GeV,
$\sigma^{Ztc} \gsim 0.2$ fb is still possible for $ 0.02 \lsim (\sin\tilde
{\alpha})^2 \lsim 0.28$ and $ 0.75 \lsim (\sin\tilde {\alpha})^2 \lsim 0.98$.

We thus conclude that at an NLC running at $\sqrt s = 500$ GeV and a yearly
integrated luminosity of ${\cal L} \gsim 10^2$ [fb]$^{-1}$ we can expect
several tens and up to hundred such $Ztc$ raw events
for $200~{\rm GeV} \lsim m_h \lsim 350$  GeV
(the number depends on $\tilde {\alpha}$ but is insensitive to  $m_H$).
However, unlike the $\nu \nu tc$
and the $eetc$ signals which form a relatively clean signature (especially at
higher energies, i.e.\  $\sqrt s \gsim 1$ TeV, where there is practically
no competing 
process that can produce a pair of $t \bar c$), the $Z t \bar c$ final state
may suffer from severe background problems if scalar FC interactions are
indeed present. For example, assuming that a $t \bar c$ pair can be detected with
some efficiency factor, still, the production rate of a pair of ${\cal H}A$
via $e^+e^- \to Z \to {\cal H} A$ followed by the decays $A \to t \bar c$ and ${\cal
H} \to f \bar f$ (recall that ${\cal H}=h~or~H$ and $f$ stands for a fermion)
may well overwhelm that of $e^+e^- \to Z \to Z t \bar c$.

\section{The Rare Top Decays $t \to W^+W^- c$, $t \to ZZc$}

Finally we wish to discuss the two rare decays $t \to W^+W^-c$ and $t \to ZZc$. The latter
being possible only if $m_t>2m_Z+m_c$ (which is still allowed by the data).
Within the SM these decay channels are vanishingly small. For the first one,
$t \to W^+W^-c$, even though a tree-level decay in the SM (i.e., the tree-level
diagram is the same as the one depicted in Fig.~1a without the
electron-neutrino fermionic lines), suffers from the same severe CKM
suppression which appears in the subprocess $W^+W^- \to t \bar c$ considered
before. Typically, one finds ${\rm Br}(t \to W^+W^-c) \approx 10^{-13}-10^{-12}$
for $160~ {\rm GeV} \lsim m_t \lsim 200 ~{\rm GeV}$~\cite{jenkins,atwood}.
For the second decay $t \to ZZc$, the branching ratio is even smaller since
it occurs only at one loop and in addition it is also GIM suppressed.

The situation is completely different in Model~III where both decay modes can
occur at the tree-level through the FC Higgs exchange of Fig.~1b
(without the leptonic lines) and the CKM factors are absent. These decays are
thus related to the fusion reactions, $WW,~ZZ \to \bar t c$, by crossing
symmetry. Therefore, in terms of the hard cross-section given in (\ref{vvhtc}):

\def\qwe{\zeta}

\begin{equation}
\Gamma_{VV} \equiv \Gamma(t \to V V c)= { m_t^3 \over 32 N_c \pi^2}
\int_{4 \qwe_V^2}^{(1- \qwe_c )^2} \!\!\!\!\!\! dz ~
z (z - 4\qwe_V^2)  \!\!\! \!\!\! \sum_{h_{V^1},h_{V^2}}
\left. {\hat {\sigma}}_V \right|_{\hat s = m_t^2 z} \label{ttovvc}~.
\end{equation}
The scaled branching-ratio $ {\rm Br}(t \to
W^+W^-c)/ \lambda ^2$
is given in Fig.~11 as a function of the light Higgs mass and for $m_t=170,180$
and 190 GeV\null. Also, in Table~1 we present the branching-ratios for both
$t \to W^+W^-c$ and $t \to ZZc$ where we focus on the range $m_t - 25~{\rm
GeV} <m_h< m_t + 25$ GeV (keeping $m_h>2m_W$). We  see that $ {\rm Br}(t \to
W^+W^-c)/ \lambda ^2$ is largest for $ 2m_W \lsim m_h \lsim m_t$ and drops
rapidly when $m_h<2m_W$ or $m_h>200$ GeV\null. 
The reason is that 
when $m_h < 2m_W$ or $m_h>m_t$, the decay $t \to W^+W^-c$ is a genuine 3-body
decay. 
Thus, it suffers a suppression factor $\sim {\rm Br}(t\to W^+W^-c)/{\rm
Br} (t\to hc)$ compared to the essentially 2-body case, $t \to hc$,
which is relevant for the window, $2m_W \lsim m_h \lsim m_t$. $ {\rm Br}(t \to 
W^+W^-c)/ \lambda ^2$ is typically a few times
$10^{-8}$ for $m_h \gsim m_t$ and can reach $ \sim 10^{-6}$ in the $m_h \lsim
2m_W$ region. For a wide range of $m_h$, i.e.\ from about 50 GeV to about
300 GeV, ${\rm Br}(t \to W^+W^-c)/ \lambda
^2$ is 3--4 orders of magnitude larger than the SM prediction.

For optimal values of $m_h$, lying in the very narrow window, $2m_W
\lsim m_h \lsim m_t$, we find that $ {\rm Br}(t \to
W^+W^-c)/ \lambda ^2$ can reach the $10^{-5}$--$10^{-4}$ level. In this region
the $t$-quark decays to an on-shell Higgs boson followed by the decay $h \to W^+W^-$. 
Note that the process $t\to c h$
studied in~\cite{savage} is related to the reaction $t\to
W^+W^-c$ under discussion here. In the region $2m_W \lsim m_h \lsim m_t$ 
the decay width satisfies $\Gamma_{WW}
\approx \Gamma(t \to ch) \times {\rm Br}(h \to W^+W^-)$. Note, however, 
that the analytical results of \cite{savage} correspond to the choice $\tilde\alpha \to 0$ and 
in this
special case Higgs decays to $WW$, $ZZ$ are suppressed at tree level even when
$m_h>2m_W$. In the present paper we use the more generic value
$\tilde\alpha=\pi/4$ in which case $h\to WW$ becomes the dominant
$h$ decay.

Concerning $t \to ZZc$, the branching ratio is typically $\sim 10^{-5}$
for $2m_Z +m_c < m_t < 200$ GeV if again $m_h$ lies in the very narrow window
$2m_Z \lsim m_h \lsim m_t$. Also, 
both decays are very sensitive to $m_t$. In Fig.~12 we have plotted $
{\rm Br}(t \to W^+W^-c)/ \lambda ^2$ and $ {\rm Br}(t \to
ZZc)/ \lambda ^2$ as a function of $m_t$ holding fixed the mass of the heavy
Higgs boson at $m_H=1$ TeV and taking $m_h=170$ and 185 GeV\null. We see that a
$\sim 10$ GeV shift in $m_t$ can easily generate an order of magnitude change
in the  branching ratios. For some possible values of $m_h$ in the range $150~{\rm GeV} \lsim m_h
\lsim 200~{\rm GeV}$ it can even generate a change of several orders of
magnitude.

\section{Summary and outlook}

In this paper, we have emphasized the importance
of searching for the FC reactions $e^+e^- \to t \bar c \nu_e {\bar {\nu}}_e$,
$e^+e^- \to t \bar c e^+ e^-$ and $e^+e^- \to Z t \bar c$ in a high
energy $e^+e^-$
collider. These reactions are very sensitive indicators of physics
beyond the SM with new FC couplings of the top quark. As an
illustrative example we have considered the consequences of extending
the scalar sector of the SM with a second scalar doublet such that new
FC couplings occur at the tree-level. At $\sqrt s= 500$ GeV the
production rates for the $Z t \bar c$ and $t \bar c \nu_e {\bar {\nu}}_e$ final
states are comparable (several tens of raw events are expected). However, for c.m.\
energies at the TeV level and above, we found that within a large
portion of the parameter space of the FC 2HDM, i.e.\ Model~III, in a
one year of running with a yearly integrated luminosity of ${\cal L} \gsim 100$--500
[fb]$^{-1}$, these new FC
couplings may give rise to several hundreds and up to a few thousands $t \bar c
\nu_e {\bar {\nu}}_e$ events and tens to hundreds 
of
$t \bar c e^+ e^-$ events in
the NLC. This will unambiguously indicate the existence of new physics.

We have shown that the comparison between $\sigma^{\nu \nu tc}$ and the
``normal'' $\sigma^{\nu \nu tt}$ comes out favorable in these models.
The $t \bar c$ final state involved, is rather distinctive and, therefore,
serious background problems for either the $t \bar
c \nu_e {\bar {\nu}}_e$ or the $t \bar c e^+ e^-$ signatures are not anticipated.
Moreover,
from the experimental point of view, it should be emphasized that although
$\sigma^{eetc}$ is found to be one order of magnitude smaller then $\sigma^{\nu
\nu tc}$, the $t \bar c e^+e^-$ signature may be 
easier to detect as it does not have the missing energy associated
with the two neutrinos in the $t \bar c \nu_e {\bar {\nu}}_e$ final state.
 Also, at $\sqrt s \gsim 1$ TeV, the 
$t \bar c \nu_e {\bar {\nu}}_e$ and $t \bar
c e^+ e^-$ signatures are to some extent unique, as 
other simple FC $s$-channel
processes like $e^+e^- \to Z \to t \bar c$, $e^+e^- \to Z{\cal H} \to Z t \bar
c$ and $e^+e^- \to A {\cal H} \to t \bar t c \bar c, t \bar c f \bar f$ 
tend to drop as $1/s$ and are therefore expected to yield much smaller
production rates at an $e^+e^-$ collider with $\sqrt{s} \gsim1$ TeV\null.

We have also examined the two rare top decays $t \to W^+W^-c$ and $t \to ZZc$.
We found that, within Model~III, the  branching ratios are many orders
of magnitudes bigger then the SM ones. However,  detection of such exotic
signatures 
may not be possible at the NLC as it is expected to produce $\sim {\rm few} \times 10^4$ $t
\bar t$ pairs. However, if nature provides us with a scalar particle, $h$,
with mass in the range $150 ~{\rm GeV} \lsim m_h \lsim 200 ~{\rm GeV}$ and
with FC couplings to $tc$, then the LHC, which will be capable of producing
$10^7-10^8$ $t \bar t$ pairs, will be able to detect those rare signatures
of top decays. 

We wish to end with the following remarks and outlook:

\begin{itemize}
\item Note that in our previous work, \cite{hep9703221}, we have used $m_H=750$ GeV while here we have set the heavy Higgs mass to be $m_H=1$ TeV. No significant difference between the two choices is observed.  

\item It is most likely that the Higgs particles, if at all present,
will have been discovered by the time the NLC starts its first
run. If indeed such a particle is detected with a mass of a few hundreds GeV,
it will be extremely important to investigate the
reactions $e^+e^- \to t \bar c \nu_e {\bar {\nu}}_e$ and
$e^+e^- \to t \bar c e^+ e^-$ in the NLC as they may serve as strong
evidence for the existence of a nonminimal scalar sector with FC
scalar couplings to fermions. In addition, since supersymmetry strongly
disfavors an $h$ heavier than $ \sim 150$ GeV, the
detection of a Higgs particle above this limit would drive the study of general
extended scalar sector, not of a supersymmetric origin,
and, in turn, this should encourage the study of FC
effects such as the ones studied in this paper.

\item The large FC effects in  $e^+e^- \to t \bar c \nu_e {\bar {\nu}}_e$
and $e^+e^- \to t \bar c e^+ e^-$ described above
may serve as a ``yardstick'' for other,
possibly large, FC effects in those same reactions. In this sense, a
model independent analysis of the reactions
$e^+e^- \to t \bar c \nu_e {\bar {\nu}}_e$ and $e^+e^- \to t \bar c e^+ e^-$
can be very useful. This can proceed by either incorporating explicit
phenomenological FC vertices of $Ztc,WWtc,ZZtc$ etc., or by considering
new effective couplings (possibly right-handed) of the W boson to the
top and a down-type quark which will affect
Fig.~1a \cite{modelindep}. Note that the effects of an effective $Ztc$ coupling,
if at all measurable, will be directly probed in the reaction
$e^+e^- \to Z \to t \bar c$ whose cross section is larger by a factor of
$\sim (\alpha/\pi)^2$ ($\alpha$ being the fine structure constant) than the
one for
$t \bar c \nu_e {\bar {\nu}}_e$ through $WW$ fusion. Therefore, if a
vanishing production rate for $e^+e^- \to Z \to t \bar c$ is measured
in a NLC with a c.m.\ energy around $\sqrt s=500$ GeV, then the
possibility of a significant $Ztc$ coupling will be basically eliminated.

\item The cross-sections for $e^+e^- \to t \bar c \nu_e {\bar {\nu}}_e$ and
$e^+e^- \to t \bar c e^+ e^-$ grow with the c.m.\ energy of the colliding
fermions. Therefore, an analogous study, for the LHC, of production of $tc$
pairs via $VV$ fusion  may be even more interesting. However, note that in
the LHC, these type of reactions are likely to suffer from much worse
background problems.
\end{itemize}

We will refer to some of these points in a later work.

\section*{Acknowledgments}

We thank David Atwood, 
Keisuke Fujii,
George Hou, Mark Sher and Daniel Wyler for discussions.
We acknowledge partial support from U.S. Israel B.S.F. (G.E. and A.S.) and from
the U.S. DOE contract numbers DE-AC02-76CH00016(BNL), DE-FG03-94ER40837(UCR). G.E. thanks the Israel Science Foundation and the Fund for the Promotion of Research at the Technion for partial support.
\bigskip 

{\it Note Added. }   
 After completion of this manuscript, which is an 
extension of our previous work \cite{hep9703221}, we became aware of a 
very recent work \cite{hep9708228} 
where (among other things) an exact calculation for the 
reaction $e^+ e^- \rightarrow t \bar c \nu_e\bar \nu_e$ is reported. 
The difference
with the effective vector boson approximation used here appears to be
at the order of $10\%$ in the range $200~{\rm GeV} < m_h < 400~{\rm GeV}$ and $1~{\rm TeV} < \sqrt s < 2~{\rm TeV}$. For $m_H \gsim 400$ GeV and $1~{\rm TeV} < \sqrt s < 2~{\rm TeV}$ the difference can be at the order of $30\%$ or so. In general the difference diminishes as $\sqrt s$ decreases.

\pagebreak

\begin{table}
\caption[entry]{ The scaled branching ratios ${\rm Br}(t \to W^+W^-c) /
\lambda^2$ and ${\rm Br}(t \to ZZc) / \lambda^2$ in units of $10^{-6}$ for
$m_H=1$ TeV, $\tilde{\alpha}=\pi/4$ and for various values of $m_t$ and
$m_h$.  The values of $m_t$ and $m_h$ are given in GeV.}
\bigskip
\begin{center}
\begin{tabular}{|l|c|c|c|c|c|c|}
\hline
& \multicolumn{3}{c|}{${\rm Br}(t \to W^+W^-c) / \lambda^2 \times 10^6$} &
\multicolumn{3}{c|}{${\rm Br}(t \to ZZc) / \lambda^2 \times 10^6$} \\
\cline{1-7}
\protect~\footnotesize$\Downarrow m_t$&
\protect~\footnotesize$m_h=175$ &
\protect~\footnotesize$m_h=185$ & \protect~\footnotesize$m_h=195$ &
\protect~\footnotesize$m_h=175$ &  \protect~\footnotesize$m_h=185$ &
\protect~\footnotesize$m_h=195$ \\
\hline
\cline{1-7}
170 & $4.74 \times 10^{-2}$  & $1.15 \times 10^{-2}$  & $4.93 \times
10^{-2}$  & /  & /  & /  \\
\cline{1-7}
175 & 0.411  & $5.71 \times 10^{-2}$  & $2.22 \times 10^{-2}$ & /  & /  & /
\\
\cline{1-7}
180 & 34.9  & 0.202  & $6.68 \times 10^{-2}$  & / & / & /  \\
\cline{1-7}
185 & 112  & 0.792  & 0.167  & $6.97 \times 10^{-4}$  & $9.88 \times
10^{-3}$  & $2.64 \times 10^{-4}$  \\
\cline{1-7}
190  & 216 & 26.0  & 0.398  & $3.03 \times 10^{-2}$  & 8.69  & $2.61
\times 10^{-2}$  \\
\cline{1-7}
195  & 336 & 82.4  & 1.15  & 0.121  & 28.8  & 0.313  \\
\cline{1-7}
200 & 466  & 158  & 20.7 & 0.282  & 55.9  & 12.8  \\
\hline
\end{tabular}
\end{center}
\end{table}

\pagebreak

\begin{center}
{\bf Figure Captions}
\end{center}

\begin{description}

\item{Fig. 1:} (a) The Standard Model diagram for $e^+e^- \to t \bar c \nu_e
{\bar {\nu}}_e$; (b) Diagrams for $e^+e^- \to t \bar c \nu_e {\bar {\nu}}_e
(e^+e^-)$ in Model~III\null.

\item{Fig. 2:} The cross-section $\sigma (e^+e^- \to t \bar c \nu_e {\bar
{\nu}}_e +  \bar t c \nu_e {\bar {\nu}}_e)$ in units of $\lambda^2$
as a function of $m_h$ for $\sqrt s=0.5,1,1.5$ and 2 TeV\null. $\tilde\alpha=\pi/4$ and we have set $\lambda=1$ in the width $\Gamma_{\cal H}$. 

\item{Fig. 3:} The cross-section $\sigma (e^+e^- \to t \bar c \nu_e {\bar
{\nu}}_e +  \bar t c \nu_e {\bar {\nu}}_e)$ in units of $\lambda^2$
as a function of $(\sin \tilde {\alpha})^2$ for $\sqrt s=1$ TeV, $m_h=250$
GeV and $m_H=250,1000$ GeV\null. $\lambda$ as in Fig.~2.

\item{Fig. 4:} The ratio $R^{\nu \nu t c}\left[ \equiv \frac{\sigma (e^+e^-
\to t\bar c\nu_e\bar\nu_e + \bar tc\nu_e \bar\nu_e)}{\sigma (e^+e^- \to \gamma\to
\mu^+\mu^-)} \right]$ for $m_H=1$ TeV, as
a function of $m_h$ for $\sqrt s=0.5,1,1.5$ and 2 TeV\null. $\lambda$ and $\tilde\alpha$ as in Fig.~2.

\item{Fig. 5:} The ratio $R^{\nu \nu t c}$ for $m_H=1$ TeV, as
a function of $\sqrt s$ for $m_h=250,350$ and 450 GeV\null. $\lambda$ and $\tilde\alpha$ as in Fig.~2. See also caption
to Fig.~4.

\item{Fig. 6:} The ratio $R^{tc/tt} \left[ \equiv \frac{\sigma (e^+e^- \to \nu_e\bar
\nu_e t\bar c+\nu_e \bar\nu_e \bar tc)}{\sigma(e^+e^- \to \nu_e\bar \nu_e
t\bar t)} \right]$ for $m_H=1$ TeV, as a
function of $m_h$ for $\sqrt s=0.5,1,1.5$ and 2 TeV\null. $\lambda$ and $\tilde\alpha$ as in Fig.~2.

\item{Fig. 7:} The Feynman Diagram for $f \bar {f'} \to t \bar c V$ in
Model~III. For $e^+e^- \to t \bar c Z$, $V=Z$, $f=e^-$ and $\bar {f'}=e^+$.

\item{Fig. 8:} The cross-section $\sigma (e^+e^- \to t \bar c Z + \bar t c Z)$
in units of $\lambda^2$ as a function of $m_h$ for $\sqrt s=0.5,1,1.5$ and 2
TeV\null. $\lambda$ and $\tilde\alpha$ as in Fig.~2.

\item{Fig. 9:} The cross-section $\sigma (e^+e^- \to t \bar c Z + \bar t c Z)$
in units of $\lambda^2$ as a function of $\sqrt s$ for $m_h=200,250,300,350$
and 400 GeV\null. $\lambda$ and $\tilde\alpha$ as in Fig.~2.

\item{Fig. 10:} The cross-section $\sigma (e^+e^- \to t \bar c Z + \bar t
c Z)$ in units of $\lambda^2$ as a function of $(\sin \tilde {\alpha})^2$
for $\sqrt s=1$ TeV, $m_h=250$ GeV and $m_H=250,1000$ GeV\null. $\lambda$ as in Fig.~2.

\item{Fig. 11:} The scaled branching ratio, $Br(t\to W^+W^-c)/\lambda^2$
as a function of $m_h$ for various values of $m_t$. $\lambda$ and $\tilde\alpha$ as in Fig.~2.

\item{Fig. 12:} The scaled branching ratios, $Br(t\to W^+W^-c)/\lambda^2$ and
$Br(t\to ZZc)/\lambda^2$ as a function of $m_t$ for $m_H=1$ TeV and $m_h=170$
and 185 GeV\null. $\lambda$ and $\tilde\alpha$ as in Fig.~2.

\end{description}

\newpage
\pagestyle{empty}

\begin{figure}
\centering
\leavevmode
\epsfysize=250pt
\epsfbox{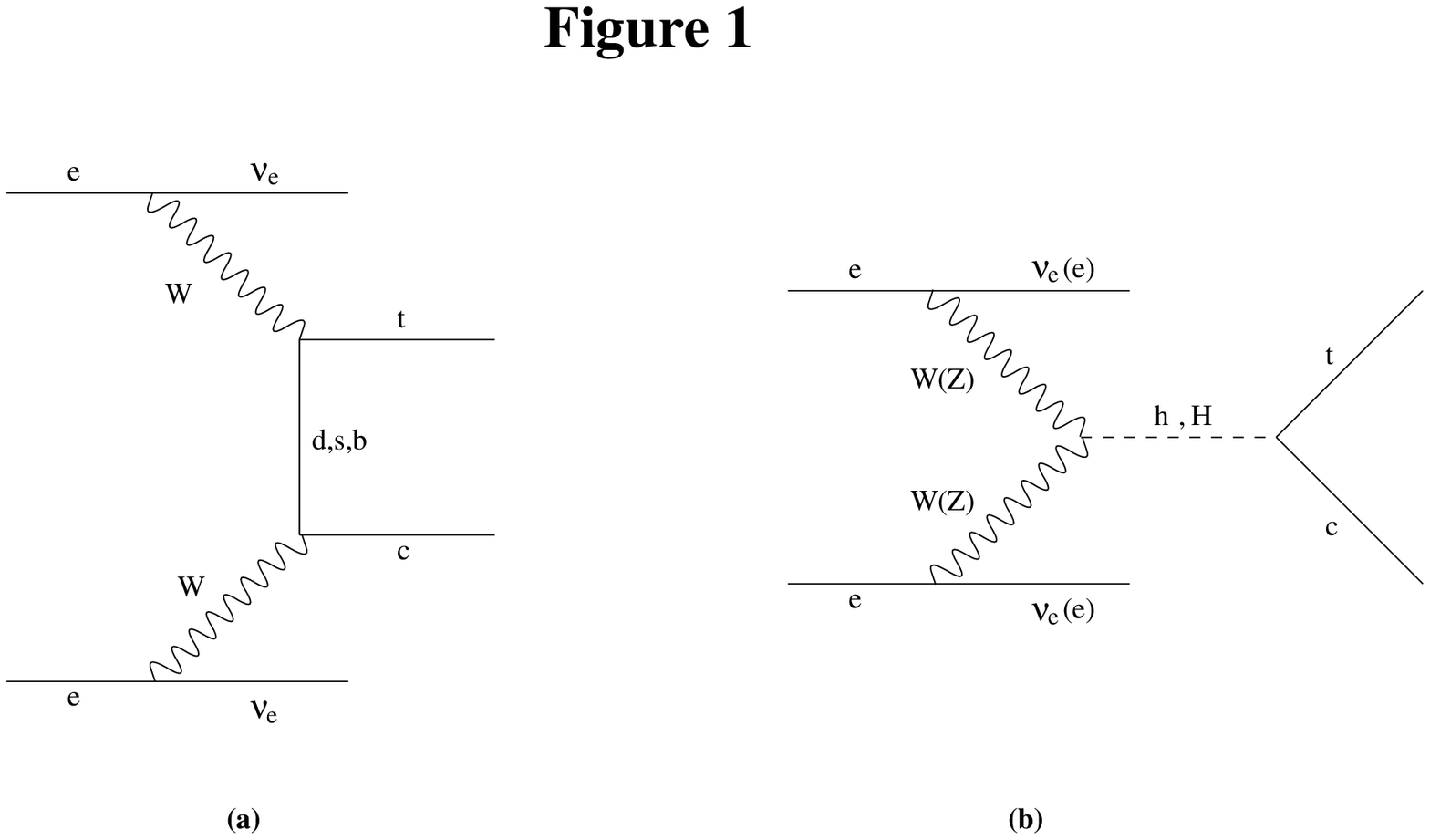}
\end{figure}

\begin{figure}
\centering
\leavevmode
\epsfysize=550pt
\epsfbox[0 0 612 792]{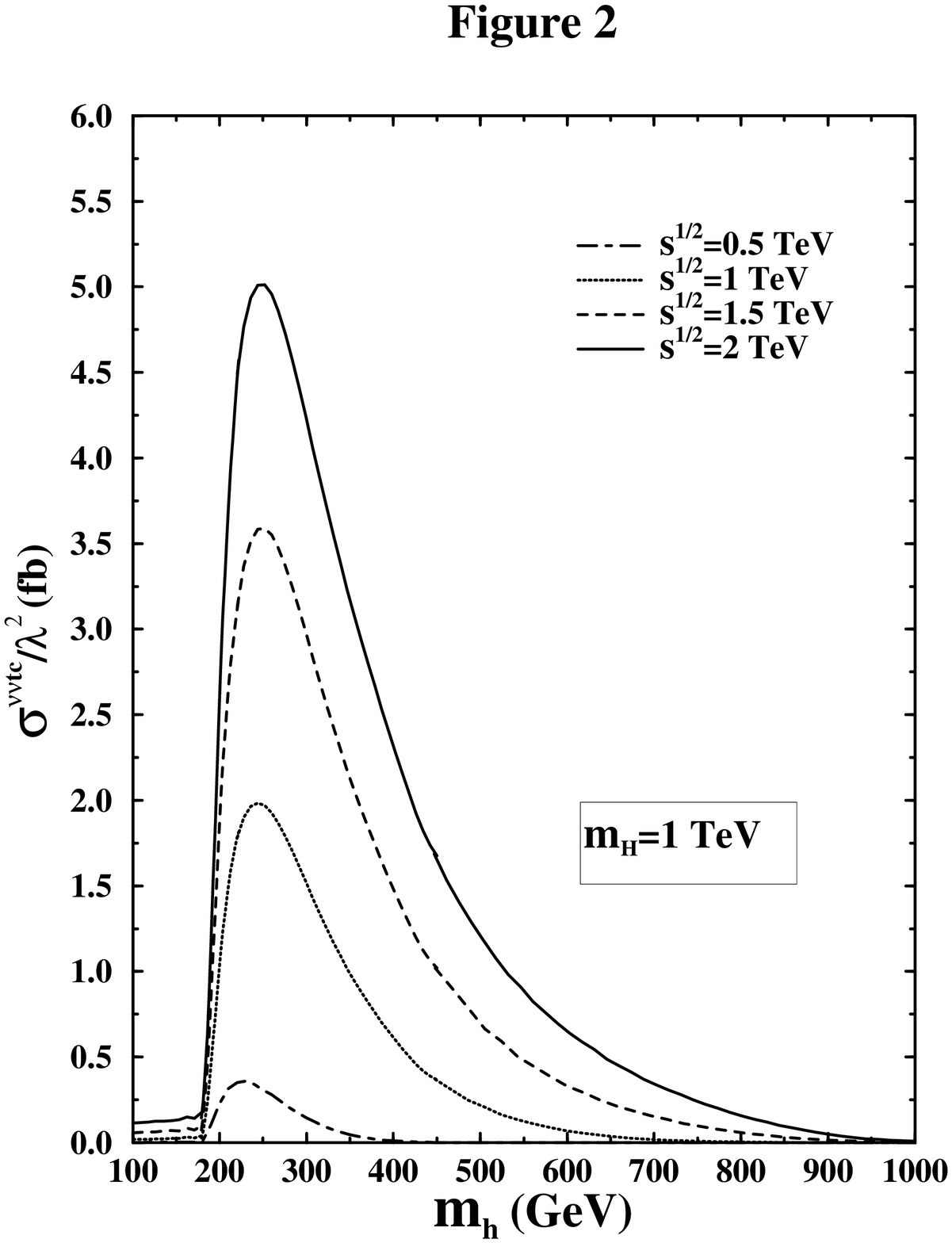}
\end{figure}

\begin{figure}
\centering
\leavevmode
\epsfysize=550pt
\epsfbox[0 0 612 792]{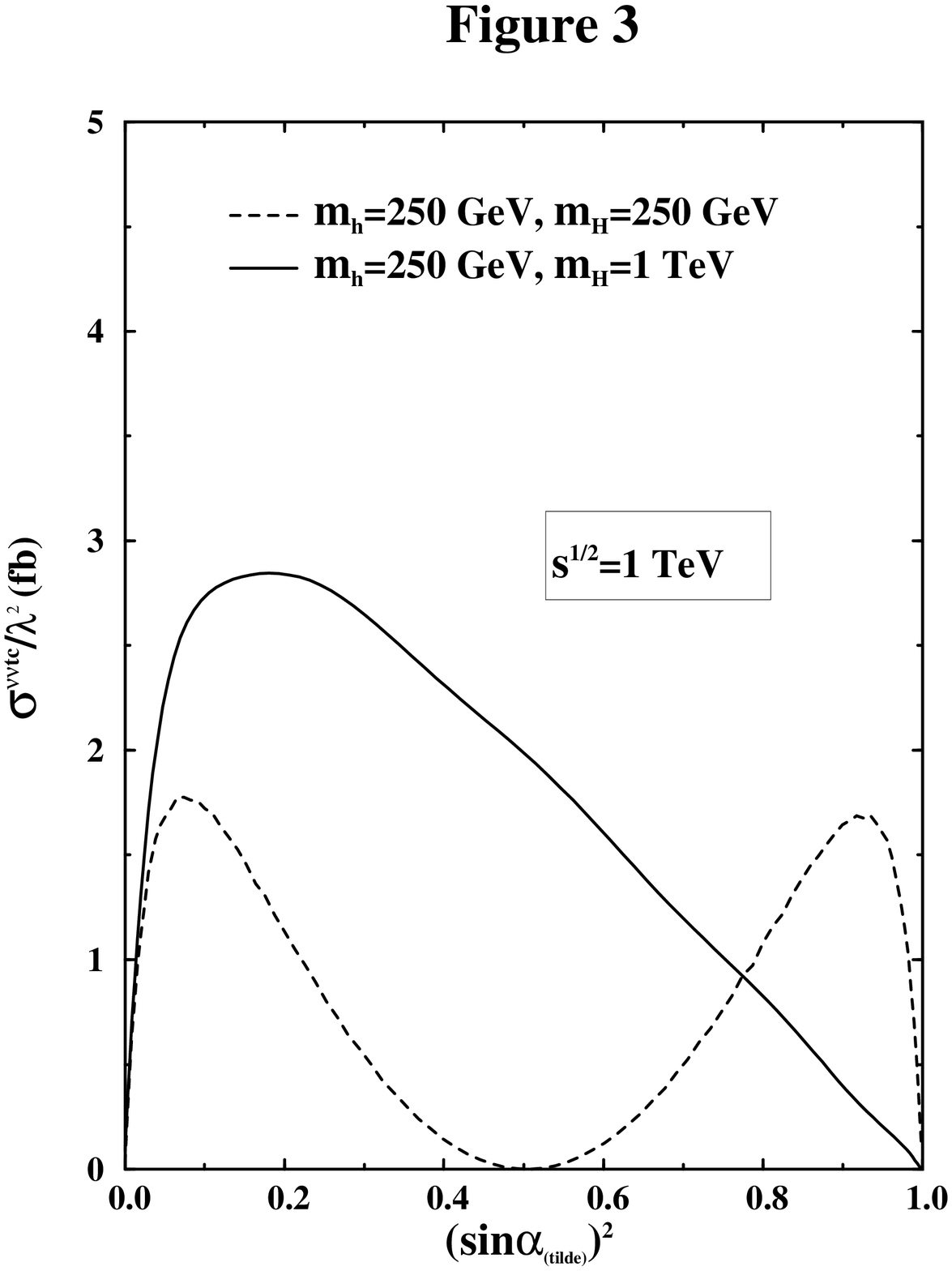}
\end{figure}

\begin{figure}
\centering
\leavevmode
\epsfysize=550pt
\epsfbox[0 0 612 792]{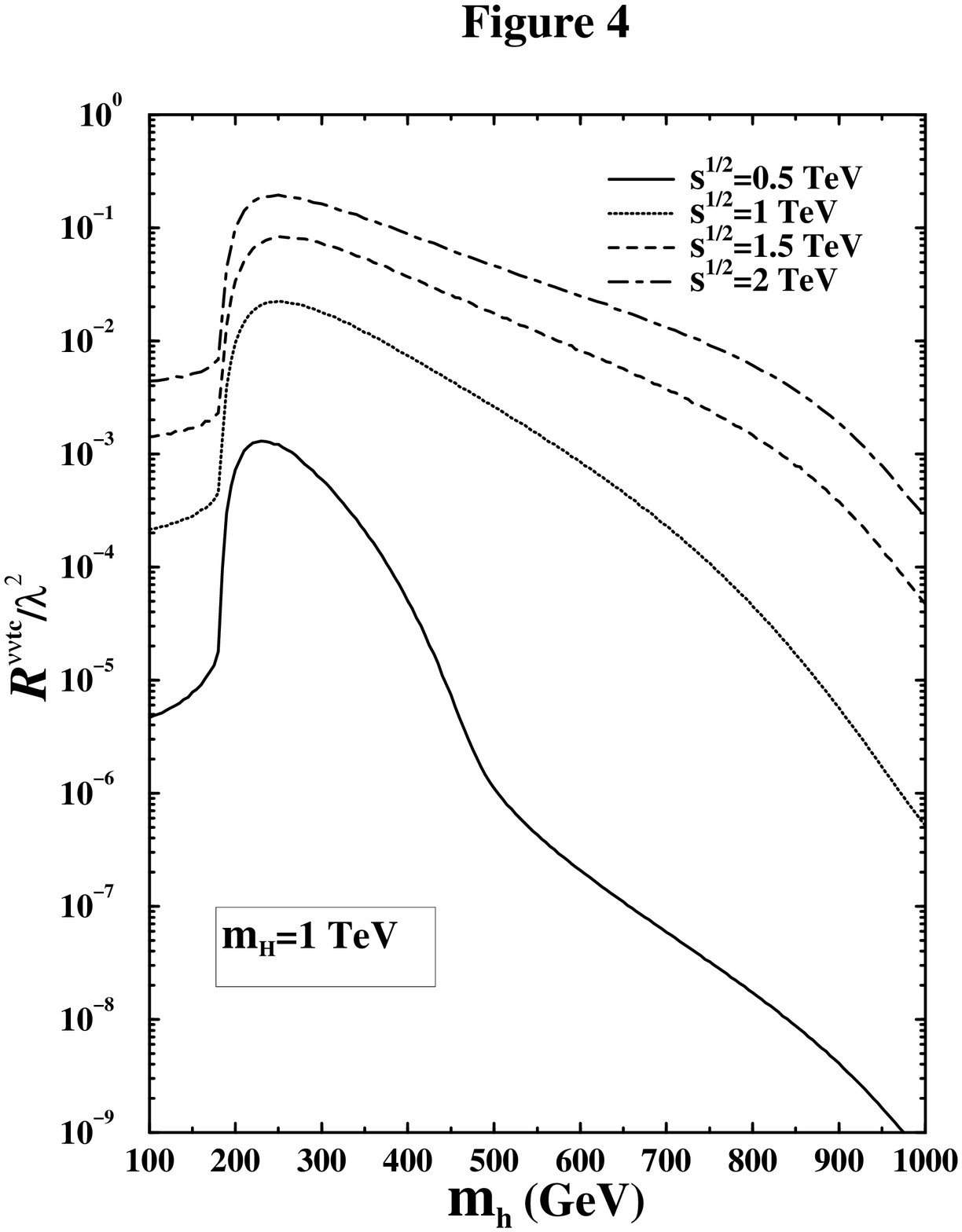}
\end{figure}

\begin{figure}
\centering
\leavevmode
\epsfysize=550pt
\epsfbox[0 0 612 792]{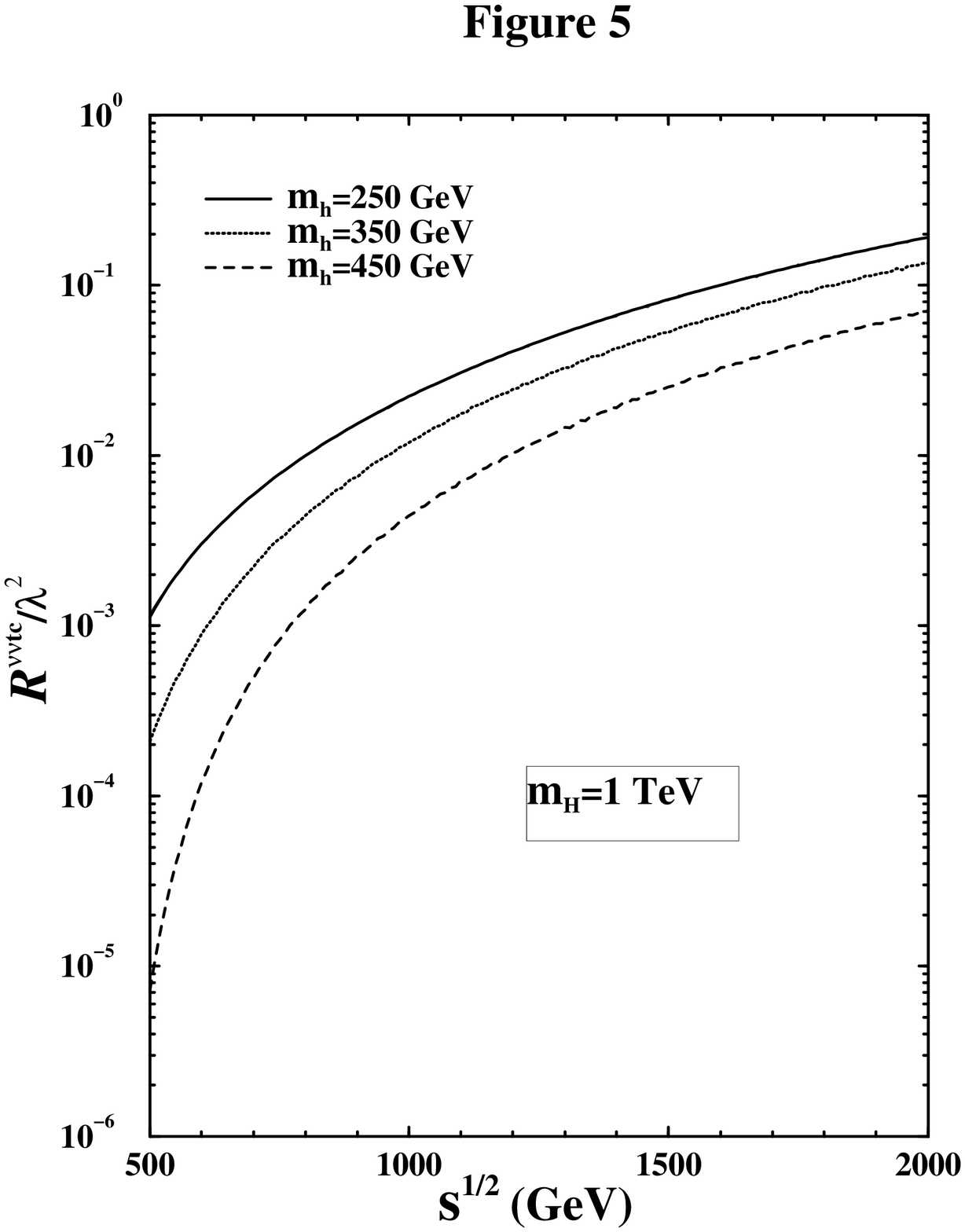}
\end{figure}

\begin{figure}
\centering
\leavevmode
\epsfysize=550pt
\epsfbox[0 0 612 792]{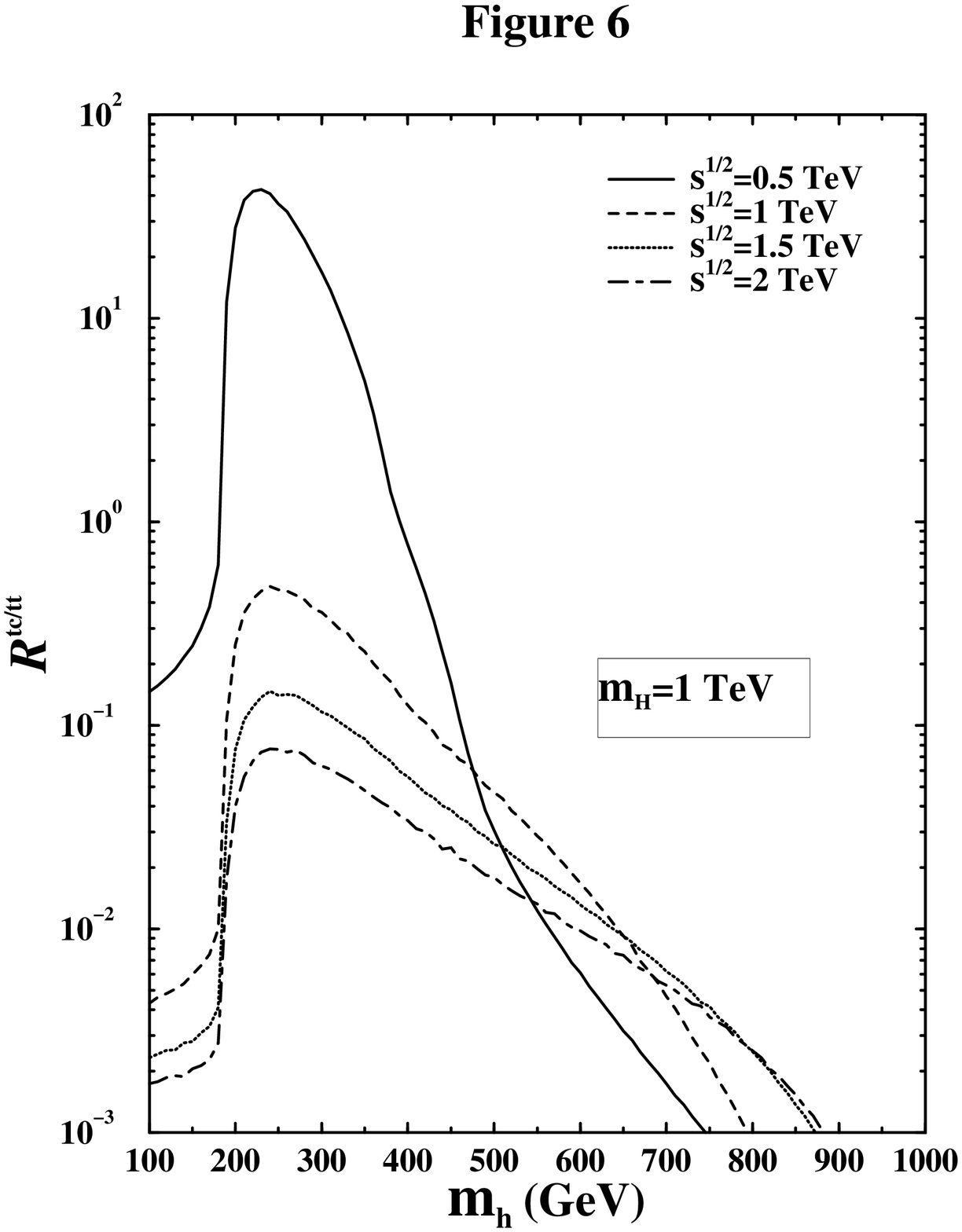}
\end{figure}

\begin{figure}
\centering
\leavevmode
\epsfysize=250pt
\epsfbox{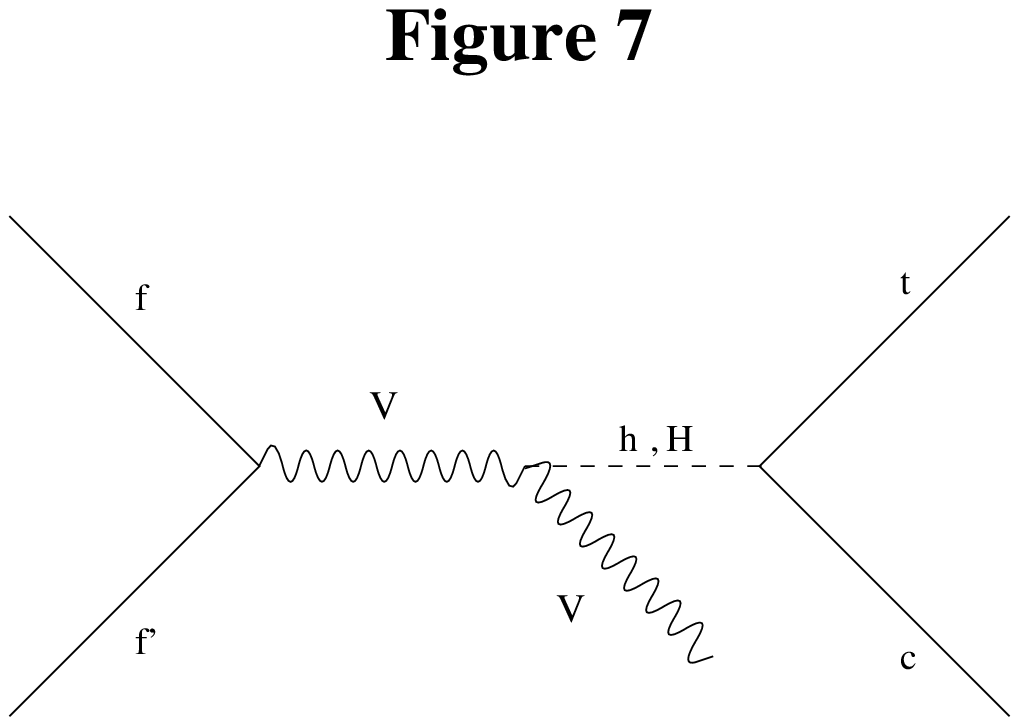}
\end{figure}

\begin{figure}
\centering
\leavevmode
\epsfysize=550pt
\epsfbox[0 0 612 792]{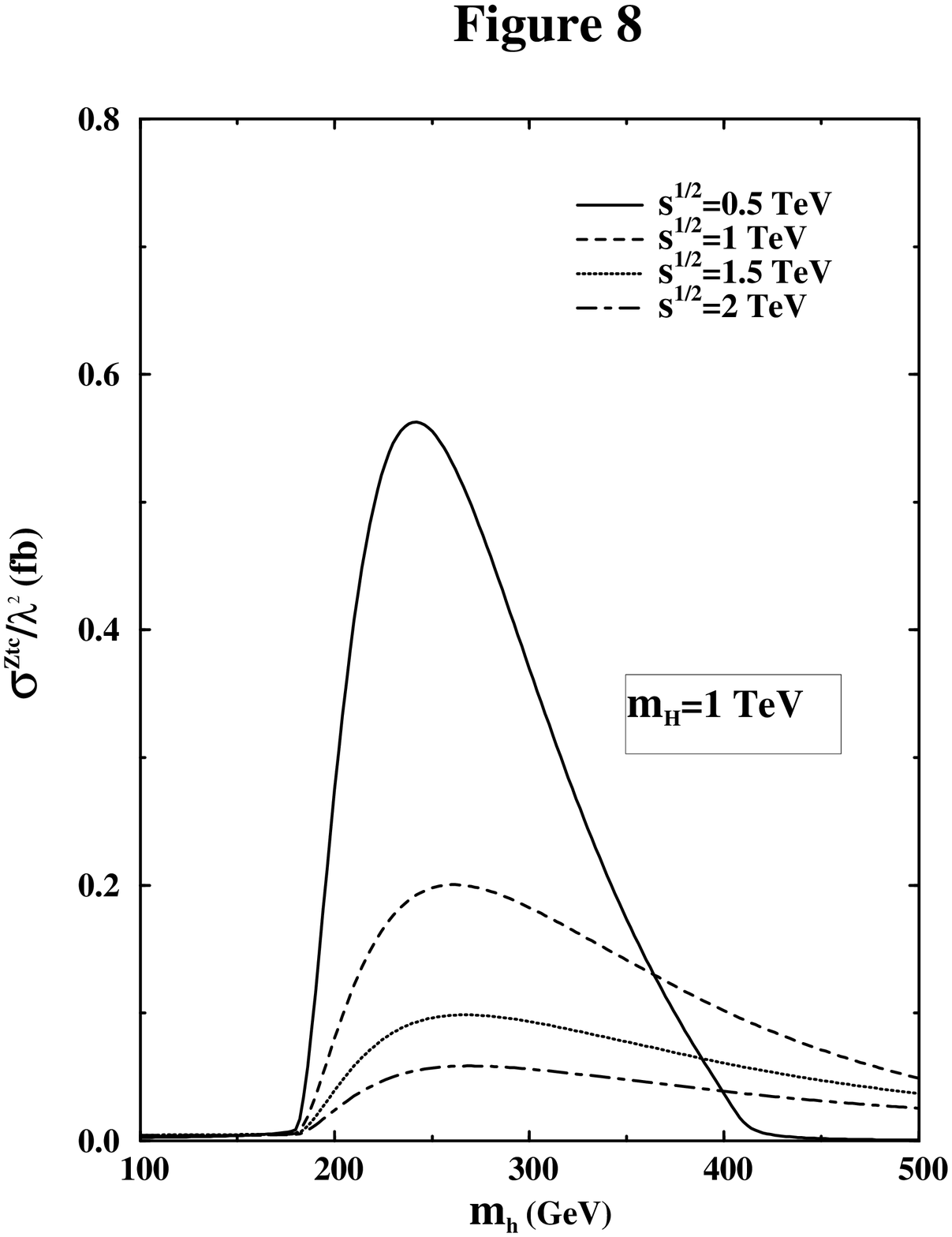}
\end{figure}

\begin{figure}
\centering
\leavevmode
\epsfysize=550pt
\epsfbox[0 0 612 792]{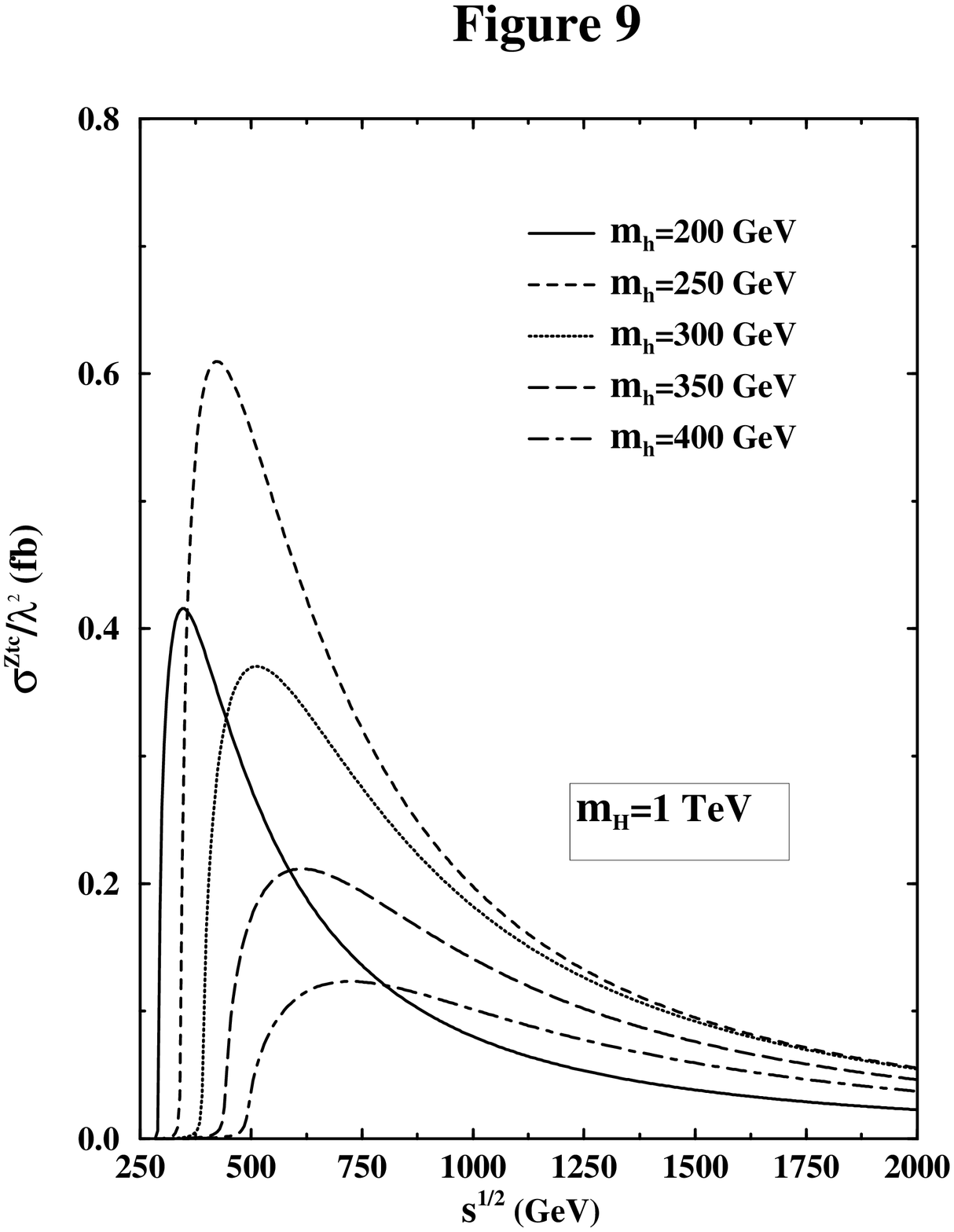}
\end{figure}

\begin{figure}
\centering
\leavevmode
\epsfysize=550pt
\epsfbox[0 0 612 792]{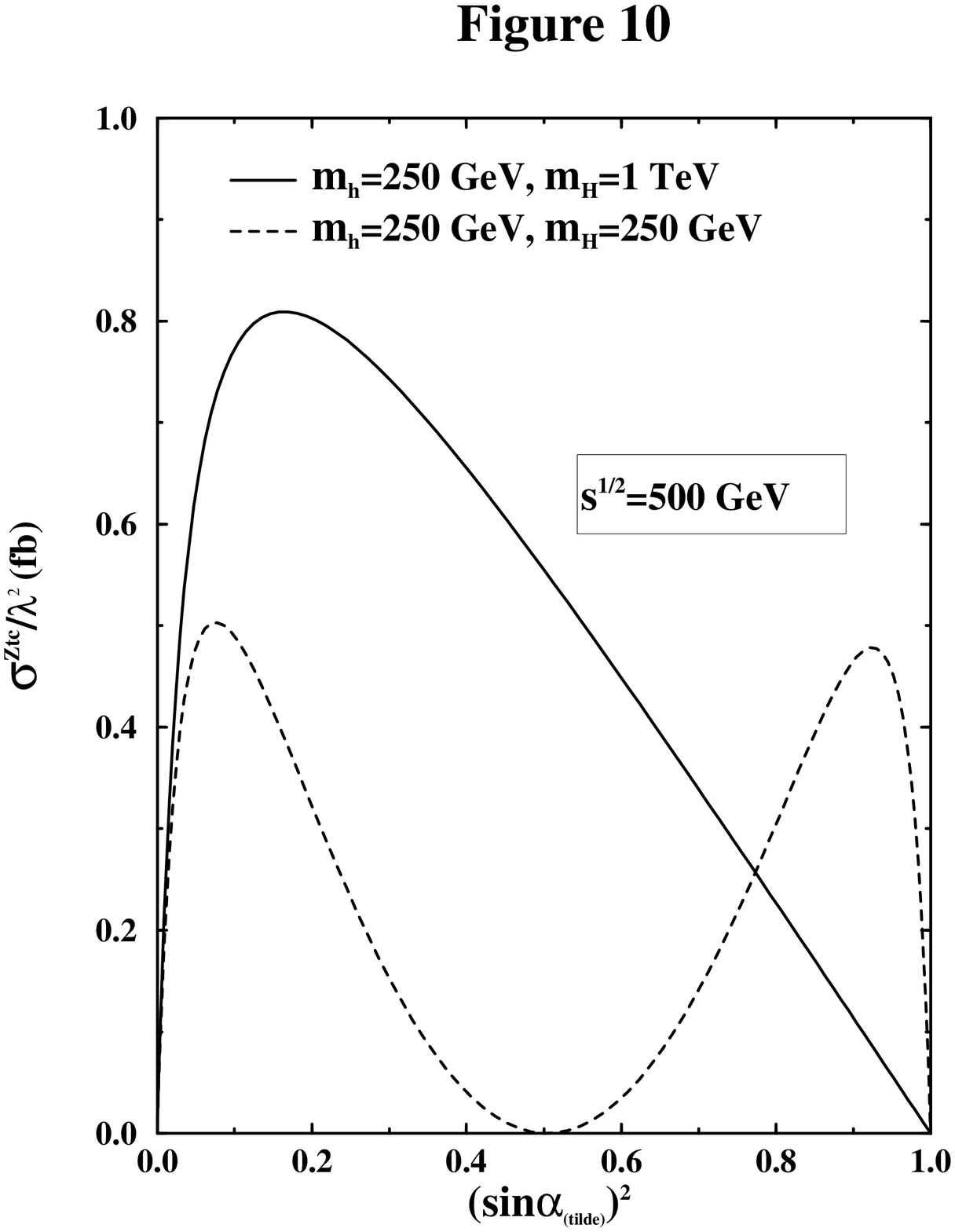}
\end{figure}

\begin{figure}
\centering
\leavevmode
\epsfysize=550pt
\epsfbox[0 0 612 792]{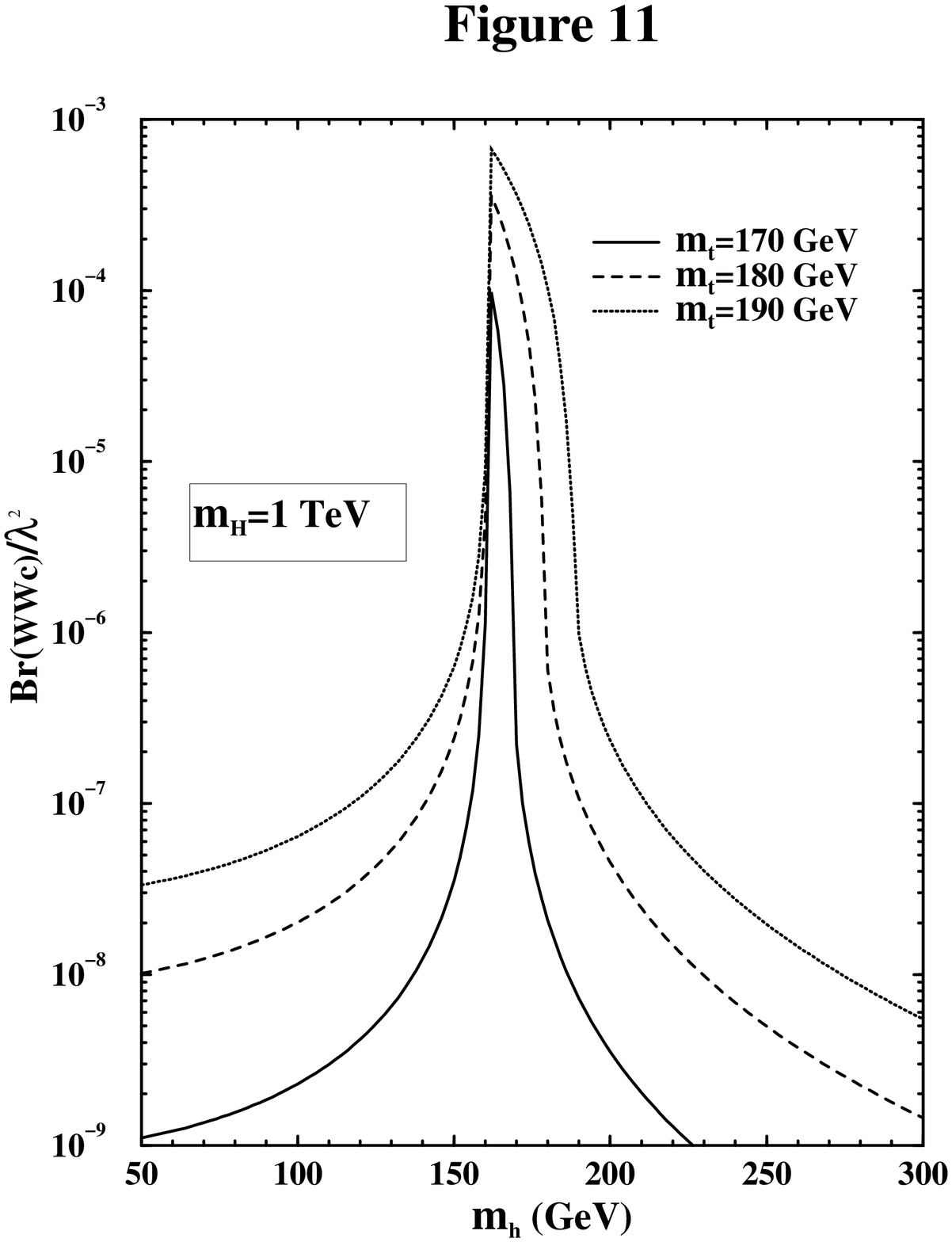}
\end{figure}

\begin{figure}
\centering
\leavevmode
\epsfysize=550pt
\epsfbox[0 0 612 792]{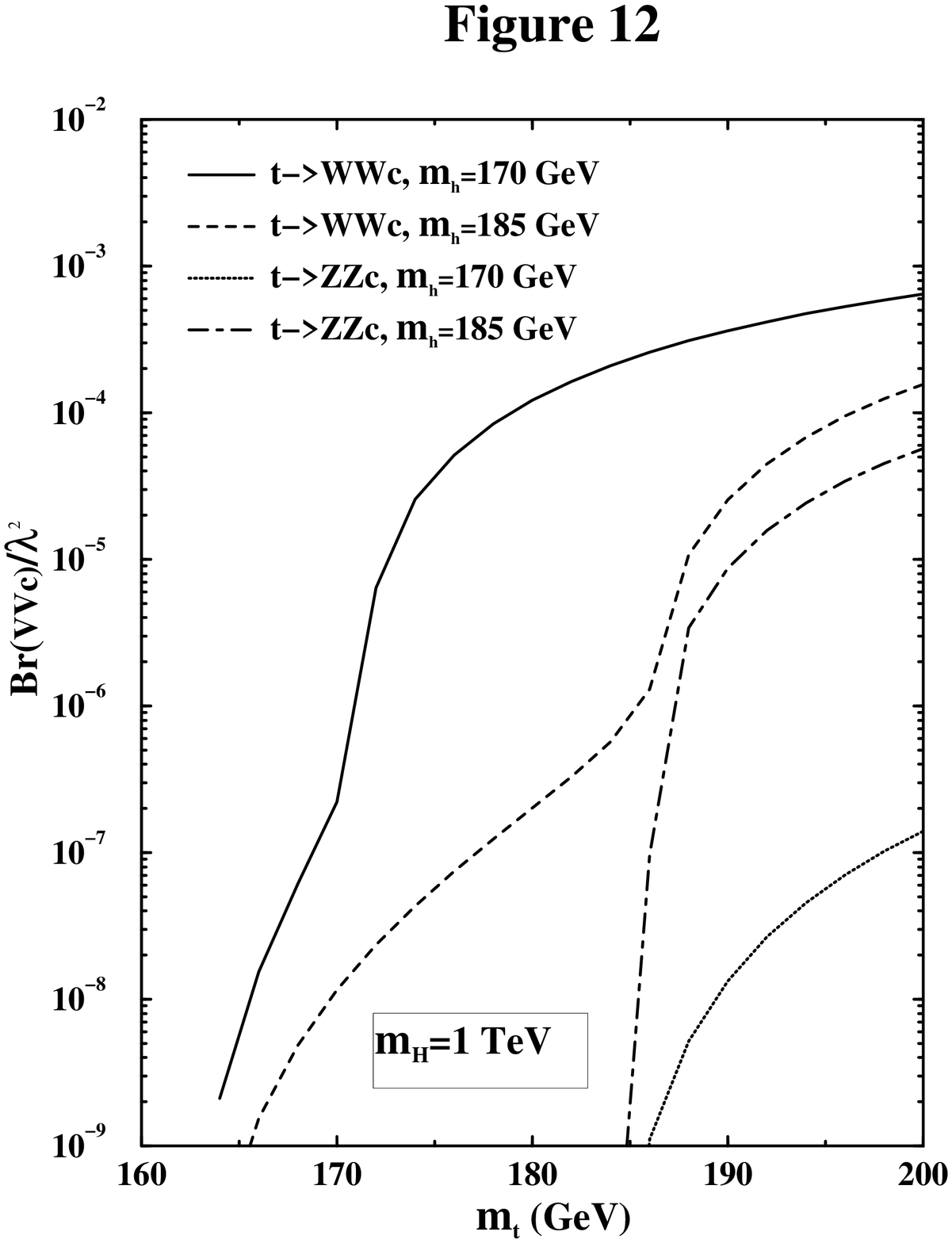}
\end{figure}

\end{document}